\RequirePackage{fix-cm}
\documentclass{article} 
\usepackage{graphicx}
\usepackage[numbers]{natbib}
\usepackage{bm}
\usepackage{multirow}
\usepackage{colortbl}
\usepackage[utf8]{inputenc}
\usepackage{epstopdf}
\usepackage{csquotes}
\MakeOuterQuote{"}
\usepackage{tikz}
\usetikzlibrary{shapes,arrows}
\usetikzlibrary{calc}
\usepackage{xurl}
\usepackage{mathptmx} 
\begin{document}
\title{The effect of Mach number on open cavity flows with thick or thin incoming boundary layers
}
\author{Marlon Sproesser Mathias\footnote{marlon.mathias@usp.br}\\
	Marcello A F Medeiros
	\\
	\small{São Carlos School of Engineering - University of São Paulo}
}
\date{}
\maketitle

\begin{abstract}
The Rossiter modes of an open cavity were studied using bi-global linear analysis, local instability analysis and nonlinear numerical simulations. Rossiter modes are normally seen only for short cavities, hence in the study, the length over depth ratio was two. We focus on the critical region, hence the Reynolds numbers based on cavity depth were close to 1000. We investigated the effect of the ratio boundary layer thickness to cavity depth, a parameter often overlooked in the literature. Increasing this ratio is destabilizing and increases the number of unstable Rossiter modes. Local instability analysis revealed that the hierarchy of unstable modes was governed by the mixing in the cavity opening. The effect of Mach number was also studied for thin and thick boundary layers. Compressibility had a very destabilizing effect at low Mach numbers. Analysis of the Rossiter mode eigenfunctions indicated that the acoustic feedback scaled to $Ma^3$, and explained the strong destabilizing effect of compressibility at low Mach numbers. At moderate Mach numbers the instability either saturated with Mach number or had an irregular dependence on it. This was associated with resonances between Rossiter modes and acoustic cavity modes. The analysis explained why this irregular dependence occurred only for higher order Rossiter modes. In this parameter region, three-dimensional modes are either stable or marginally unstable. Two-dimensional simulations were performed to evaluate how much of the nonlinear regime could be captured by the linear stability results. The instability was triggered by the $10^{-13}$ flow solver noise floor. The simulations initially agreed with linear theory, and later became nonlinearly saturated. The simulations showed that, as the flow becomes more unstable, an increasingly more complex final stage is reached. Yet, the spectra present distinct tones that are not far from linear predictions, with the thin boundary layer cases being closer to empirical predictions. The final stage, in general, was dominated by first Rossiter mode, even though the second one was the most unstable linearly. It seems this may be associated with nonlinear boundary layer thickening, which favors lower frequency in the mixing layer, or vortex pairing of the second Rossiter mode. The spectra in the final stages are well described by the mode R1 and a cascade of nonlinearly generated harmonics, with little reminiscence of the linear instability.
\end{abstract}
\section{Introduction}
\begin{figure}[h!]
 \begin{center}
 \includegraphics[width=0.8\textwidth]{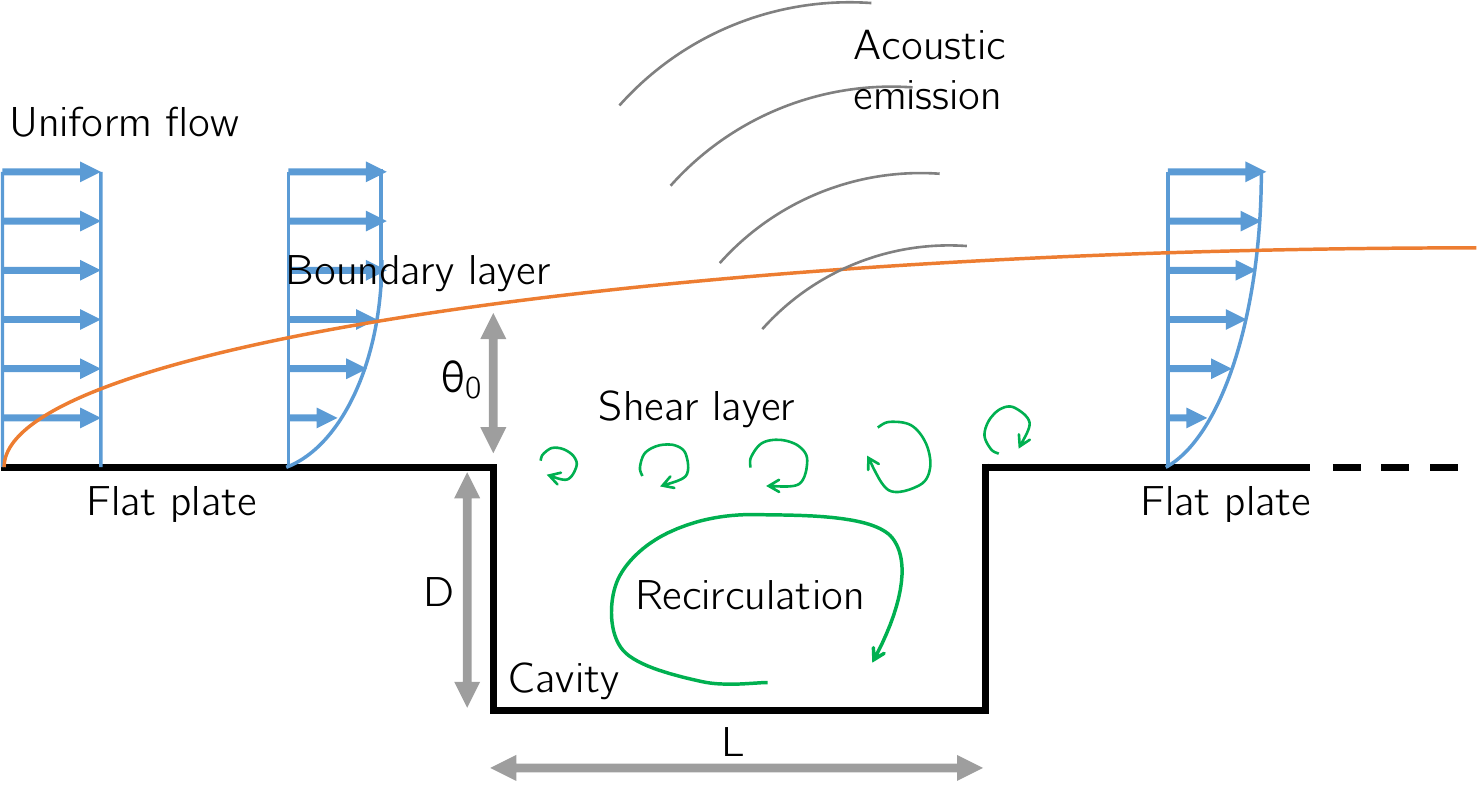}
 \end{center}
 \caption{Illustration of the open cavity flow.}
 \label{fig:cavityDescription}
\end{figure}

An open cavity is a canonical geometry that can represent several features of vehicles, ranging from very small ones, such as gaps at doors and windows, to very large ones, such as the landing gear wells of airplanes and car sun roofs. Figure~\ref{fig:cavityDescription} illustrates the flow over a cavity, and defines the parameters which, together with Reynolds and Mach numbers, govern the flow, namely, cavity depth ($D$), cavity length ($L$) and momentum thickness of the incoming boundary layer ($\theta$).

A mixing layer forms at the cavity opening and the flow circulates inside the cavity. Under certain circumstances, oscillations develop, and high levels of sound are emitted. In transonic and supersonic flows, such oscillations can greatly increase the aerodynamic drag of surfaces with cavities \cite{Mcgregor1970}. \citet{Owen1958} indicates that pressure fluctuations in the cavities can affect the aircraft structure and \citet{Dix2000b} point out they can cause structural fatigue.

Cavity flow and its noise emission is a long-standing research topic. It initiates with \citet{Krishnamurty1956}, who observed experimentally acoustic tones emanating from the flow over a two-dimensional gap. \citet{Rossiter1964} extended the study and described a two dimensional mechanism for the tone formation, illustrated in figure~\ref{fig:cavityDescription}. A shear layer initiates at the cavity leading edge giving rise the Kelvin-Helmholtz (KH) vortices. These impinge on the cavity trailing edge and produce sound which interacts with the cavity leading edge triggering new vortices. Based on this mechanism, \citet{Rossiter1964} proposed an equation which predicted the tone frequencies, and these oscillations are now called Rossiter modes. \citet{Plumblee1962} introduce another type of mode in rectangular cavities, the acoustic cavity modes, which was later extended to other geometries by \citet{Tam1976}. The frequencies of these modes can be theoretically predicted for appropriate boundary conditions. \citet{East1966} indicated the possibility of resonance between Rossiter and acoustic modes.  Experiments by \citet{Gharib1987} revealed the existence of yet another mode in the open cavity flow, the wake mode. 

More in-depth insight into the 2D instability modes of the open cavity flow were provided by computational studies. A large number of papers are dedicated to the control of cavity flow oscillations (see \citet{Rowley2006} for a review), and a smaller number dedicate more thoroughly to the flow physics. Simulations by \citet{Rowley2002} confirmed the acoustic feedback described by \citet{Rossiter1964} and the growth of KH vortices in the mixing layer. They also demonstrate that longer cavities favor the wake mode as opposed to the Rossiter mode. Simulations by \citet{Bres2008} indicated that compressibility reduced the critical Reynolds number.  \citet{Yamouni2013} used global instability analysis \cite{Theofilis2011} and observed that the Rossiter modes growth rates had several peaks as Mach number was increased. They demonstrated that the peaks corresponded to resonances between Rossiter and acoustic cavity modes. \citet{Sun2017} extended the studies of \citet{Bres2008} establishing a neutral curve as a function of Reynolds and Mach number, indicating that compressibility can be stabilizing at higher subsonic Mach numbers, a result they suggest confirms \cite{Yamouni2013}.

In this paper we focus on Rossiter modes, which, as seen above, belong to short cavities. Table~\ref{tab:parameters} compares the parametric region of this study to other present in the literature. Despite the extensive work that covered a wide range of parameters, there remains open questions. In particular there are four points that we want to address. (1) In the studies of \citet{Rowley2002,Bres2008,Sun2017} the effect of $L/\theta$ was investigated by increasing $L$ and promoting the wake mode. In their investigations of short cavities ($L/D\le 2$), $D/\theta$ was kept between 7.5 and 29.2, with most tests for $D/\theta=26.4$. \citet{Yamouni2013} covered $D/\theta$ from 26.4 to 231, but only to indicated that lower $D/\theta$ was stabilizing. Their resonance analysis was restricted to $D/\theta=231$. It appeared to us that the effect of $D/\theta$ on the Rossiter modes (short cavities) has not been sufficiently studied. The parameter $\theta$ affects the instability of the mixing layer, and, as such, has the potential to affect the Rossiter mode selection, an aspect that has not been investigated. (2) According to \citet{Yamouni2013}, which investigated very unstable conditions ($Re_D=7500$, $D/\theta=231$) the effect of Mach is as follows. The potential for the K-H instability is given at $Ma=0$. As Mach number increases there is competition between the stabilization of the mixing layer \cite{Miles1958,Pavithran1989} and the destabilizing effect of the resonance with the acoustic modes. According to \citet{Sun2017}, who investigated conditions close to critical, compressibility is destabilizing at low Mach and stabilizing as transonic conditions are approached. These are not necessarily conflicting conclusions, but it is unclear whether the resonance interactions are present in the critical region or at small $D/\theta$. (3) The noise emission, which is part of the Rossiter mechanism, can be affected by the Mach number. This could affect the overall instability mechanism, a possibility that was not investigated by \citet{Yamouni2013} nor by \citet{Sun2017}. (4) Despite very comprehensive, studies of the Rossiter modes utilized either instability analysis or numerical simulations. An investigation about how much of the fully nonlinear regime can be captured or explained by linear stability has not been performed in a systematic way.

\begin{table}[h!]
	\begin{center}
		\caption{Summary of parametric spaces covered by the literature}
		\setlength{\tabcolsep}{2pt}
		\begin{tabular}{l|ccc|ccc|ccc|ccc|ccc|ccc}
								& \multicolumn{3}{c|}{$Ma$}    & \multicolumn{3}{c|}{$Re_D$}        & \multicolumn{3}{c|}{$Re_\theta$} & \multicolumn{3}{c|}{$L/\theta$} & \multicolumn{3}{c|}{$D/\theta$} & \multicolumn{3}{c}{$L/D$} \\ 
			\hline 
			Present work        & 0.1&-&0.9 & 1000&-&1149     & 5&-&100       & 20&-&400     & 10&-&200     & \multicolumn{3}{c}{2}     \\
			\citet{Rowley2002}  & 0.2&-&0.8 & 440&-&2000      & 29.3&-&80.5   & 20.3&-&123.2 & 7.5&-&29.2   & 1&-&8   \\
			\citet{Bres2008}    & 0.1&-&0.8 & 450&-&6960      & 35&-&400      & 23.2&-&60.2  & 15&-&26.4    & 1&-&4   \\
			\citet{Yamouni2013} & 0&-&0.9   & \multicolumn{3}{c|}{7500}          & 32.5&-&284    & 34.2&-&231   & 26.4&-&231   & 1&-&2   \\
			\citet{Sun2017}     & 0.1&-&1.4 & 132&-&3900      & 5&-&150       & 52.8&-&158.4 & \multicolumn{3}{c|}{26.4}       & 2&-&6   \\
		\end{tabular} 
		\label{tab:parameters}
	\end{center}
\end{table}


As opposed to \citet{Yamouni2013}, we want to investigate conditions close to critical, which led us to lower $Re_D$ numbers, guided by \citet{Rowley2002,Bres2008,Sun2017}. Since we focus on the Rossiter modes only, a short cavity was used with a fixed $L/D=2$. After this introduction, the numerical simulation and global instability analysis tools used are described (section~\ref{sec:methods}). In section~\ref{sec:theta} we investigate the effect of $D/\theta$ on the global instability and on the Rossiter mode hierarchy. The mechanism of frequency selection is also investigated. Then, two representative scenarios of $D/\theta$ (small and large) are selected for an investigation of the effect of Mach number in section~\ref{sec:mach}. We also investigate mechanisms of Mach number destabilization and mode selection. In section~\ref{sec:nonlinear} we compare nonlinear simulation results with linear stability results and Rossiter empirical predictions; and discuss the origin of significant discrepancies. In section~\ref{sec:conclusion} we draw some conclusions.

In the manuscript we use the term "linear approximation of Rossiter mode" or, for short, "linear Rossiter mode" to refer to the solution of the bi-global stability analysis because "Rossiter mode" is already used to refer to the high amplitude limit cycle oscillations originally observed in experiments. The use of the word "linear" makes it clear which flow entity is meant.

Early experimental studies revealed the existence of much lower frequency oscillations with a definite three-dimensional structure. These oscillations were investigated extensively by \citet{Bres2008} and in the literature are referred to as centrifugal modes \cite{DeVicente2014,Meseguer2014,Citro2015}. According to \citet{Sun2017}, compressibility has only a small effect on the 3D centrifugal modes and, consequently, the 2D modes tend to dominate as the Mach number increases. As an example, \citet{Bres2008} showed that, for $D/\theta=26.4$, 3D modes become unstable only above $Re_D=1300$. They also showed that the critical Reynolds number for 2D instability is lower than that of 3D instability for Mach numbers above around 0.4 for this $D/\theta$ ratio. Based on those results, the analysis performed here is in a region of the parameter space where the Rossiter modes are substantially dominant over the centrifugal ones and for this reason it is restricted to 2D. However, the Rossiter modes are robust and have been observed as the dominant feature in many high Reynolds number experiments even for an incoming turbulent boundary layer. Moreover, the instability analysis of 2D Rossiter modes performed for very unstable conditions (high Reynolds number, for example), where 3D modes were definitely also unstable, explained several important aspects observed in experiments \cite{Yamouni2013}. Hence, even though the analysis is here limited to 2D, it may also contribute to the more general 3D flow. In fact \citet{Yamouni2013} used arguments similar to ours to justify their two dimensional approach.

\section{Methodology}
\label{sec:methods}

\subsection{Numerical Simulation}

We used an in-house Direct Numerical Solver (DNS) \cite{Martinez2016,Mathias2018}, which features structured meshes that are refined in regions of interest. A fourth-order Runge-Kutta scheme is used for time marching and fourth-order compact spectral-like finite differences are used for the spatial derivatives \cite{Lele1992}. A pencil-slab domain decomposition is used for code parallelization \cite{Li2010}. A tenth-order spatial high-frequency filter is also employed \cite{Gaitonde1998} to prevent very short wavelength spurious oscillations. Buffer zones are placed around the useful domain to attenuate undesirable open boundary condition effects such as reflections. They employ a combination of grid stretching, lower order spatial derivatives and Selective Frequency Damping (SFD) \cite{Akervik2006}. The SFD acts as a low pass temporal filter and may also be turned on in the whole domain to allow base flows to be generated faster or at unstable conditions. Appendix~\ref{sec:validationSolver} brings validation results. Further details of these methods and their implementation in our codes are given by \citet{Souza2005,Silva2010,Bergamo2015}

\subsection{Instability Analysis}

The bi-global analysis was performed by a time-stepping approach, in which the Jacobian matrix of the governing equations is not explicitly needed \cite{Theofilis2011,Gomez2015a}. The method uses the Arnoldi algorithm \cite{Arnoldi1951} which is based on Krylov subspaces. It just requires the ability to compute vector multiplications which, due to the way in which the algorithm is built, corresponds to a call to the flow numerical solver, in our case, the code described in the previous section.

The time-stepping global instability analysis can be regarded as an established procedure and the current implementation closely followed that of \citet{Chiba1998,Tezuka2006}. In summary, the method iteratively disturbs the base flow and uses the DNS to capture its response. The successive iteration involves disturbances that are orthogonal to all previous ones. The flow response is used to form a corresponding Hessemberg matrix, which is several orders of magnitude smaller than the flow's Jacobian matrix. If the number of iterations is sufficiently large, the leading eigenvalues and eigenvectors computed from this matrix are good representations of the flow modes and provide good estimates of their respective amplification rates and frequency. In our convention, the real part of the eigenvalue represents the growth rate in time, while the imaginary part represents its angular frequency. Appendix~\ref{sec:validationGlobal} provides validation results. Further details on the implementation are given by \citet{Mathias2018}.

\subsection{Base flows}

\begin{figure}[h!]
	\begin{center}
		\includegraphics[width=0.7\textwidth]{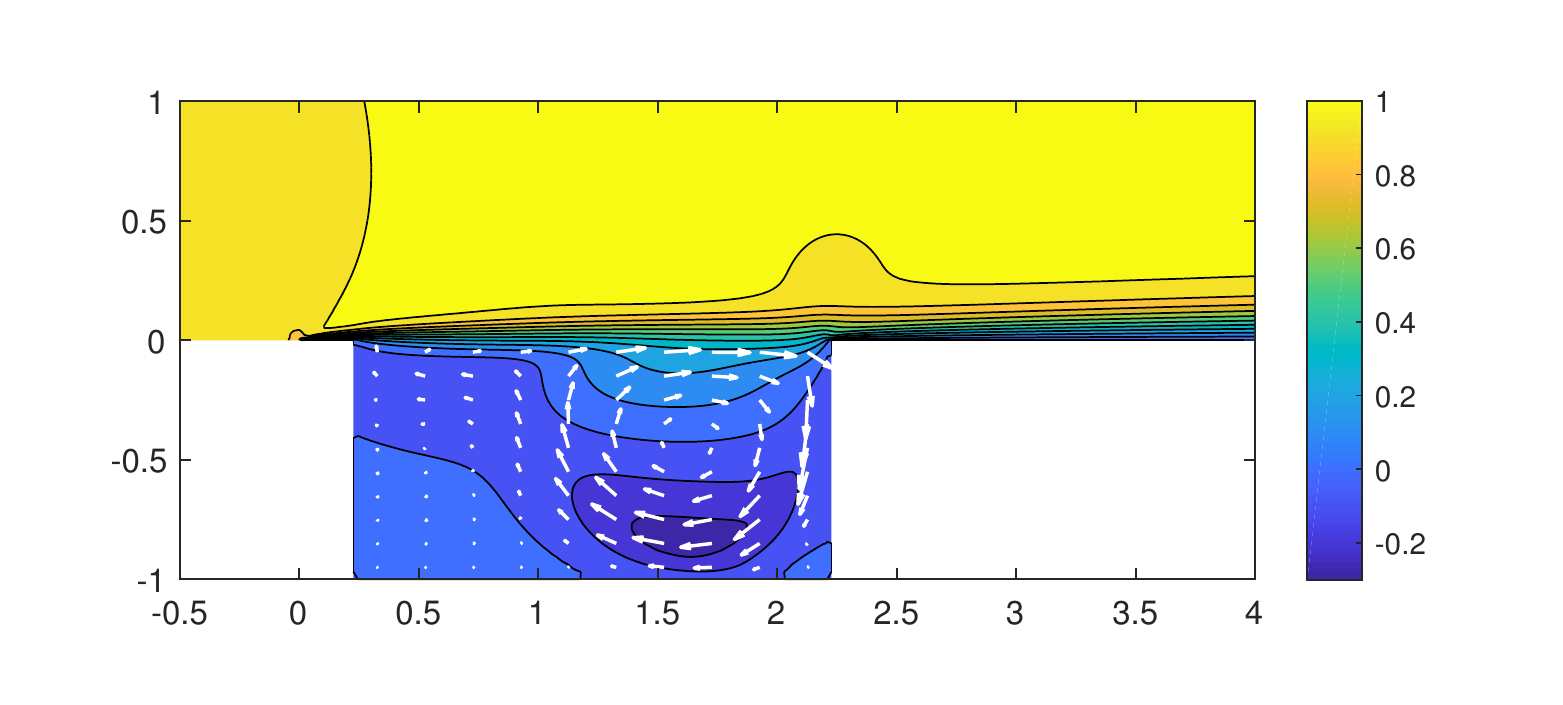}
	\end{center}
	\caption{Base flow for $Re_D=1000$, $D/\theta=100$ and $Ma=0.5$. Each contour represents a 10\% step of stream-wise velocity normalized by the free flow velocity.}
	\label{fig:baseflow100}
\end{figure}

As an example, figure~\ref{fig:baseflow100} shows the base flow for a thin boundary layer condition ($D/\theta=100$) and $Ma=0.5$. The maximum backward velocity of the flow inside the cavity is $23\%$ of the free flow velocity. The backflow increased with $D/\theta$, but reaches a saturation at this condition, and was unaffected by the Mach number. Despite the considerable backflow, the flows were not absolutely unstable, as verified in Appendix~\ref{sec:absoluteInstability}.

\section{Influence of the boundary layer thickness}
\label{sec:theta}

\subsection{Bi-global flow instability}

With the computational parameters defined and base flows obtained, the instability analysis was carried out. In the resulting eigenvalue spectra, we preliminarily selected the linear Rossiter mode from the other linear modes by comparing with the Rossiter empirical frequency prediction. After that, we looked at their eigenfunctions for confirmation.

\begin{figure}[h!]
 \begin{center}
 \includegraphics[width=0.7\textwidth]{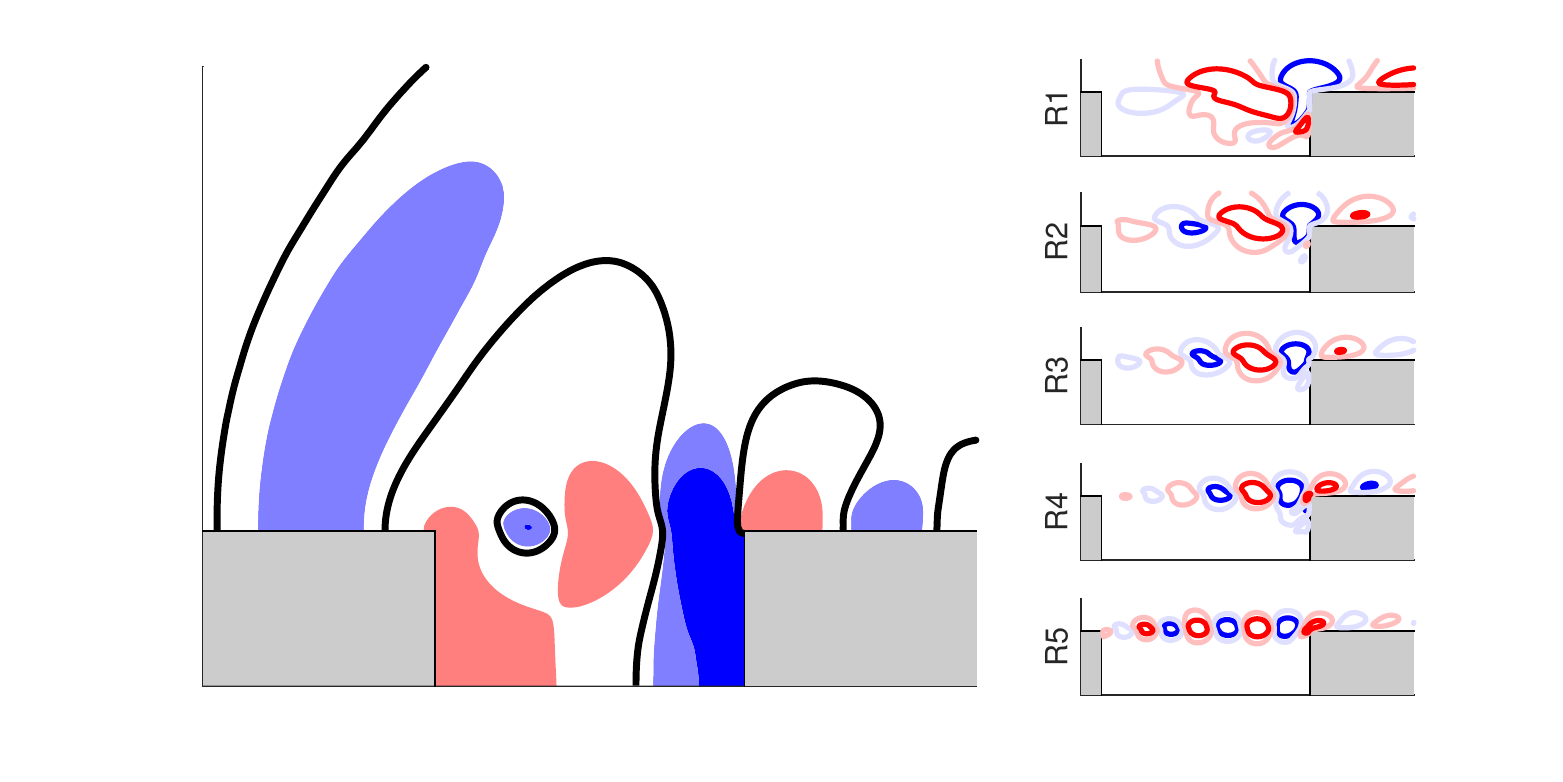}
 \end{center}
 \caption{Eigenfunctions of Rossiter mode linear approximations at $Re_D=1000$, $D/\theta=100$ and $Ma=0.5$, at an arbitrary phase. (a) Contours of pressure for mode 2. Black lines indicated level zero. (b) Contours of wall-normal velocity for modes 1 to 5.}
 \label{fig:RossiterContours}
\end{figure}

Figure~\ref{fig:RossiterContours} shows results for $D/\theta=100$, $Ma=0.5$. The left-hand side gives the pressure fluctuation contours of mode 2 at an arbitrary phase, the thick black lines represent fluctuation amplitude level zero. The right-hand side shows the wall-normal velocity contours for modes 1 to 5, which will be referred in this paper as R1 through R5. The mode number refers to the number of vortices in the mixing layer.

The Rossiter modes are described as a feedback mechanism involving four steps, namely: (1) the growth of KH vortices in the mixing layer at the cavity opening, (2) the triggering of acoustic waves at the trailing edge of the cavity by these vortices, (3) the propagation of the acoustic waves upstream and (4) the excitation of KH instability at the cavity leading edge. It is however not entirely clear how much of the complicated behavior obtained from the global instability can be explained by the phenomena just described. In particular we wanted to investigate the effect of $D/\theta$ on the Rossiter mode instability, the destabilizing effect of Mach on R1 and R2 and the complex effect of compressibility on R3 and R4.

\begin{figure}[h!]
	\begin{center}
		\includegraphics[width=0.9\textwidth]{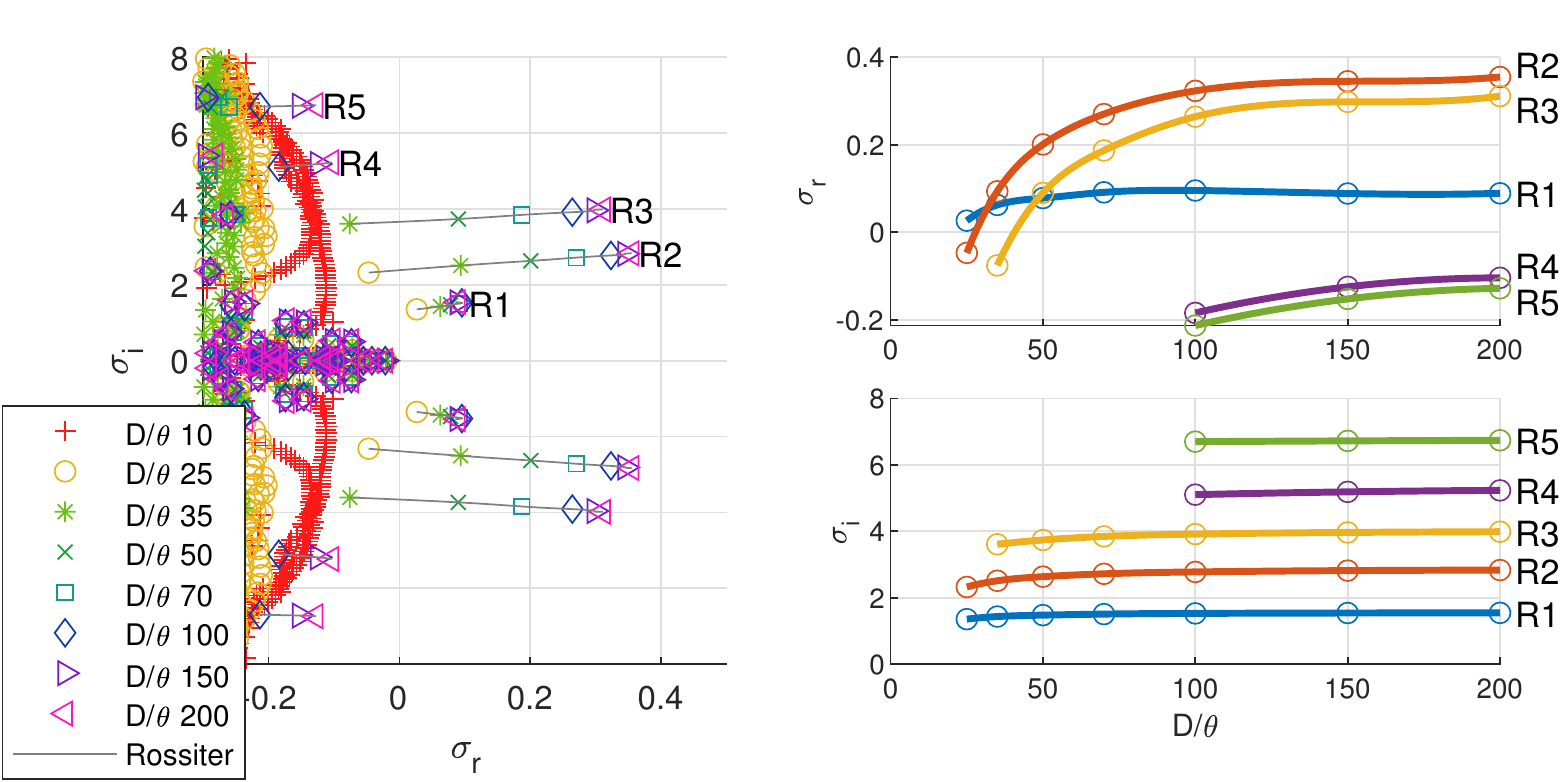}
	\end{center}
	\caption{Eigenvalues from the $D/\theta$ ratio sweep ($Re_D=1000$, $Ma=0.5$). (a) Eigenspectrum. (b) Real and imaginary parts of Rossiter mode eigenvalues.}
	\label{fig:thetaSweep}
\end{figure}

A sweep of $D/\theta$ was performed and is shown in figure~\ref{fig:thetaSweep}, the Mach number was fixed at $Ma=0.5$ and $D/\theta$ ranged between 10 and 200. In all cases, $L/D=2$ and $Re_D=1000$. Higher values of $D/\theta$ enhanced the instability. For very small $D/\theta$, only R1 is unstable and the higher the Rossiter mode the more stable it is. Rossiter modes R4 and R5 were still immersed in the mode cloud and not identifiable, and for this reason their eigenvalues are not shown. In the parameter range investigated, the growth rates increases with $D/\theta$, confirming previous findings \cite{Yamouni2013}, but reach a saturation for large $D/\theta$. Higher Rossiter modes show stronger variation of growth rates with respect to $D/\theta$ and saturate at higher values of $D/\theta$. For these reasons, at higher $D/\theta$, R2 overtakes mode R1 and becomes dominant. Afterwards, R3 also becomes more unstable than R1. The overall picture conveys the idea that as $D/\theta$ increases, higher Rossiter modes become dominant if a saturation is not reached before that. For the current parameter range only R1 and R2 dominate, but it seems that for higher $Re$ or longer cavities, higher Rossiter modes could become dominant. At the same time, a slight increase in the mode frequency was observed. This confirms previous studies which attribute this effect to a higher convective velocity of the KH vortices in the mixing layer.

\subsection{Physical mechanism of mode selection}

\citet{Rowley2002} estimated the growth of the instability waves by integrating the local spatial growth rates for the velocity profiles along the mixing layer length. The same approach was used here, but including the viscous effects which they neglected. In our calculations the compressibility effects were not considered because, as also suggested by \citet{Yamouni2013}, for $Ma=0.5$, they are still very small \cite{Ragab1989,Germanos2009}, in particular with regard to the wavenumber of the most unstable mode, which is our main concern.

\begin{figure}[h!]
	\begin{center}
		\includegraphics[width=\textwidth]{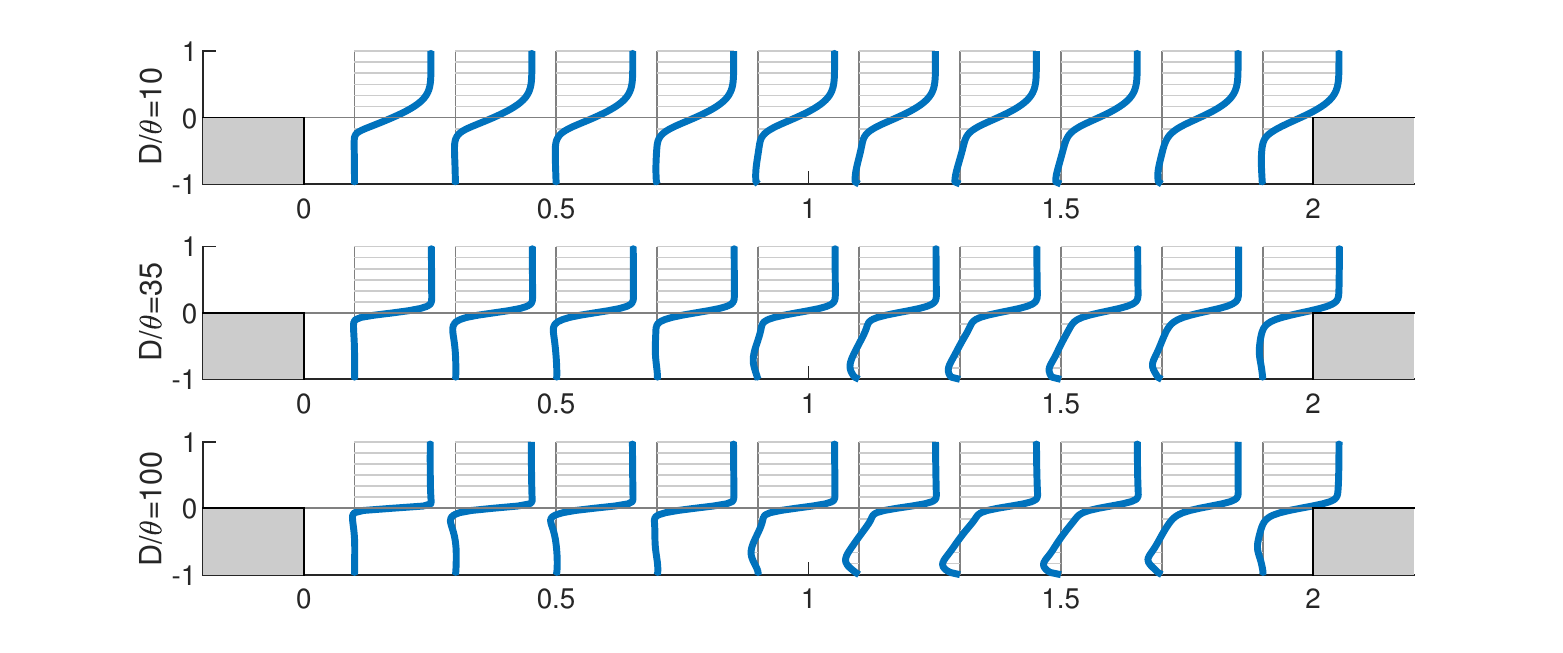}
	\end{center}
	\caption{Mixing layer profile for an incoming boundary layer thicknesses of $D/\theta=10, 35$ and $100$.}
	\label{fig:mixingLayerProfile}
\end{figure}

Figure~\ref{fig:mixingLayerProfile} shows the velocity profiles for several positions along the cavity, and for different $D/\theta$. These velocity profiles were extracted from the base flows used for the global stability analysis. Results are for $Ma=0.5$, but, within the subsonic regime, the Mach number has a negligible effect on these profiles.

\begin{figure}[h!]
	\begin{center}
		\includegraphics[width=\textwidth]{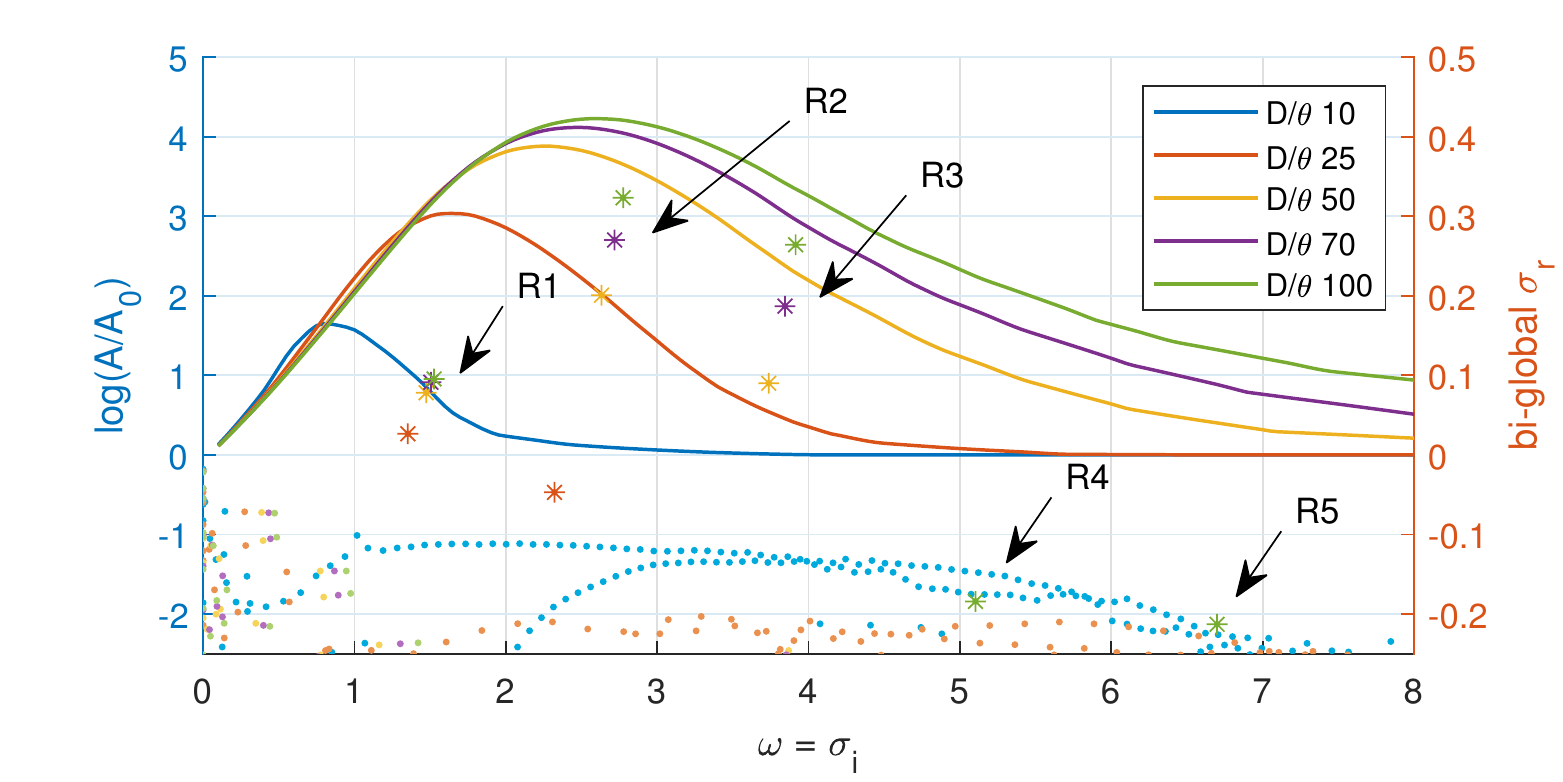}
	\end{center}
	\caption{Mixing layer integrated spatial amplification factor compared to the global instability results. The asterisks indicate the global modes eigenvalues of Rossiter modes and the dots are other eigenvalues, and refer to the scale on the right hand side of the frame.}
	\label{fig:mixingLayerAmplification}
\end{figure}

The spatial growth of each profile in the mixing layer was obtained by the Orr-Sommerfeld equation. These amplification rates were integrated along the mixing layer to obtain the total spatial growth for each frequency $\omega$, which are shown as the full lines in figure~\ref{fig:mixingLayerAmplification} for different $D/\theta$. In the analysis, the last 5\% of the cavity length were disregarded, as the parallel-flow approximation is invalid there. The picture also includes bi-global stability growth rates of the Rossiter modes R1 to R5, which are referred to the vertical scale on the right-hand side of the frame.

At $D/\theta=10$, the global analysis shows complete stability of Rossiter modes. The mixing layer is unstable only to very low frequencies and even the lowest Rossiter mode frequency is only marginally unstable in the mixing layer. At $D/\theta=25$ mode R1 is the only globally unstable, and its frequency is close to the most unstable for the mixing layer. At $D/\theta=50$, R2 is the dominant globally unstable mode followed by R1 and R3 with similar growth rates. Consistently the mixing layer predicts R2 to be very close to the most unstable mode, with R1 and R3 on each side of the instability curve maximum.

$D/\theta=70$ and $100$ lead to progressively smaller effects on the global instability results. This is consistent with the small variation of mixing layer instability results. At $D/\theta\ge50$, R1 in particular is virtually unaffected by $D/\theta$ both in the global and the local analysis. As the boundary layer becomes thinner, higher frequencies have their instability increased, which in turn allows mode R3 to become more unstable and even modes R4 and R5 to emerge from the cloud of stable modes represented by the dots in the figure.

Overall, all the major aspects of the effect of $D/\theta$ on the global instability were in perfect qualitative agreement with the integrated local analysis of the mixing layer. The hierarchy of Rossiter modes in this parameter range was determined by the mixing layer instability alone.

\section{Influence of the Mach number on thick and thin boundary layers}
\label{sec:mach}

\subsection{Bi-global flow instability}

\begin{figure}[h!]
	\begin{center}
		\includegraphics[width=0.9\textwidth]{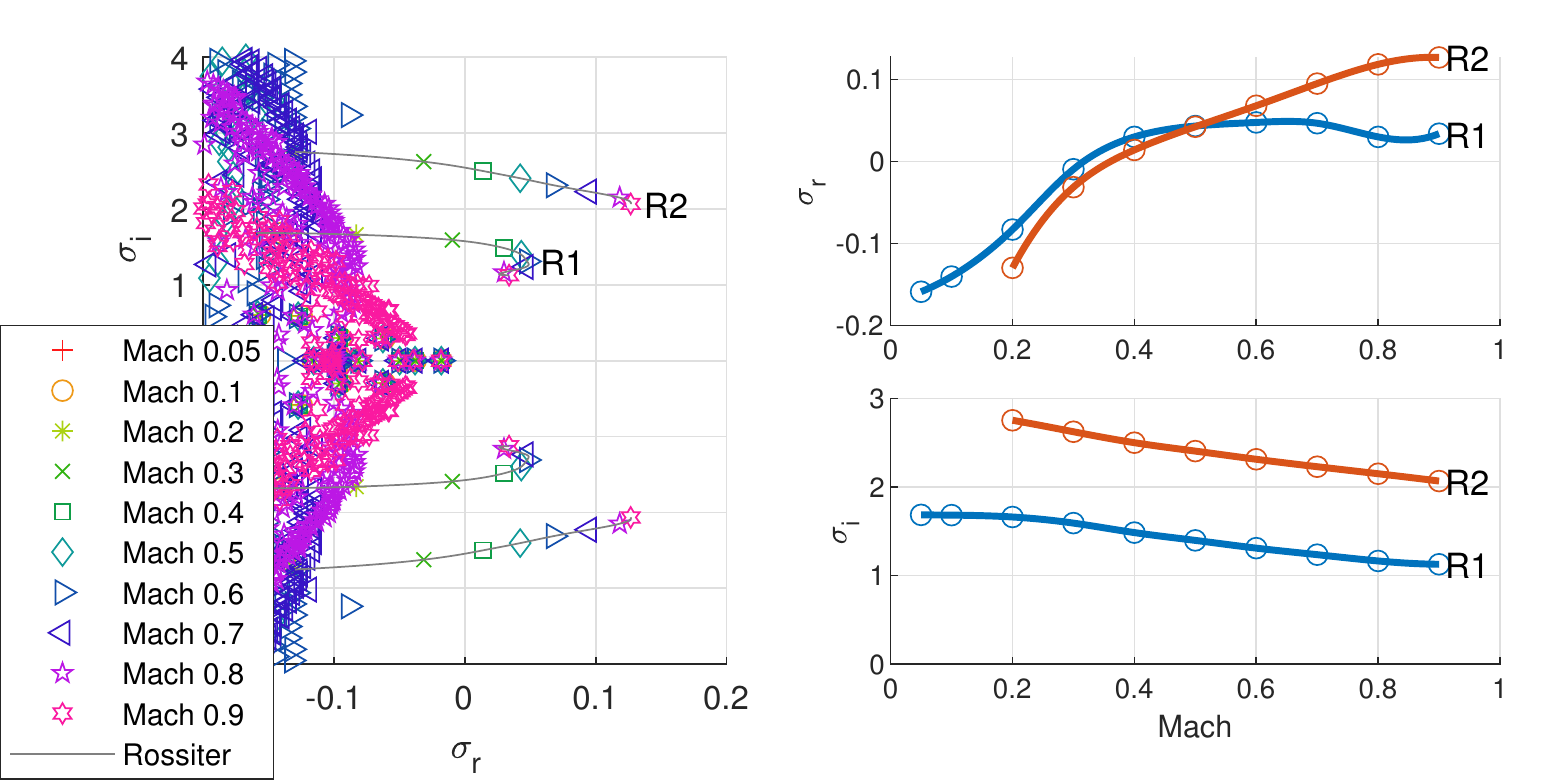}
	\end{center}
	\caption{Eigenvalues from the first Mach number sweep ($Re_D=1149$, $D/\theta=29.7$). (a) Eigenspectrum. (b) Real and imaginary parts of Rossiter mode eigenvalues.}
	\label{fig:machSweep1}
\end{figure}

\begin{figure}[h!]
	\begin{center}
		\includegraphics[width=0.9\textwidth]{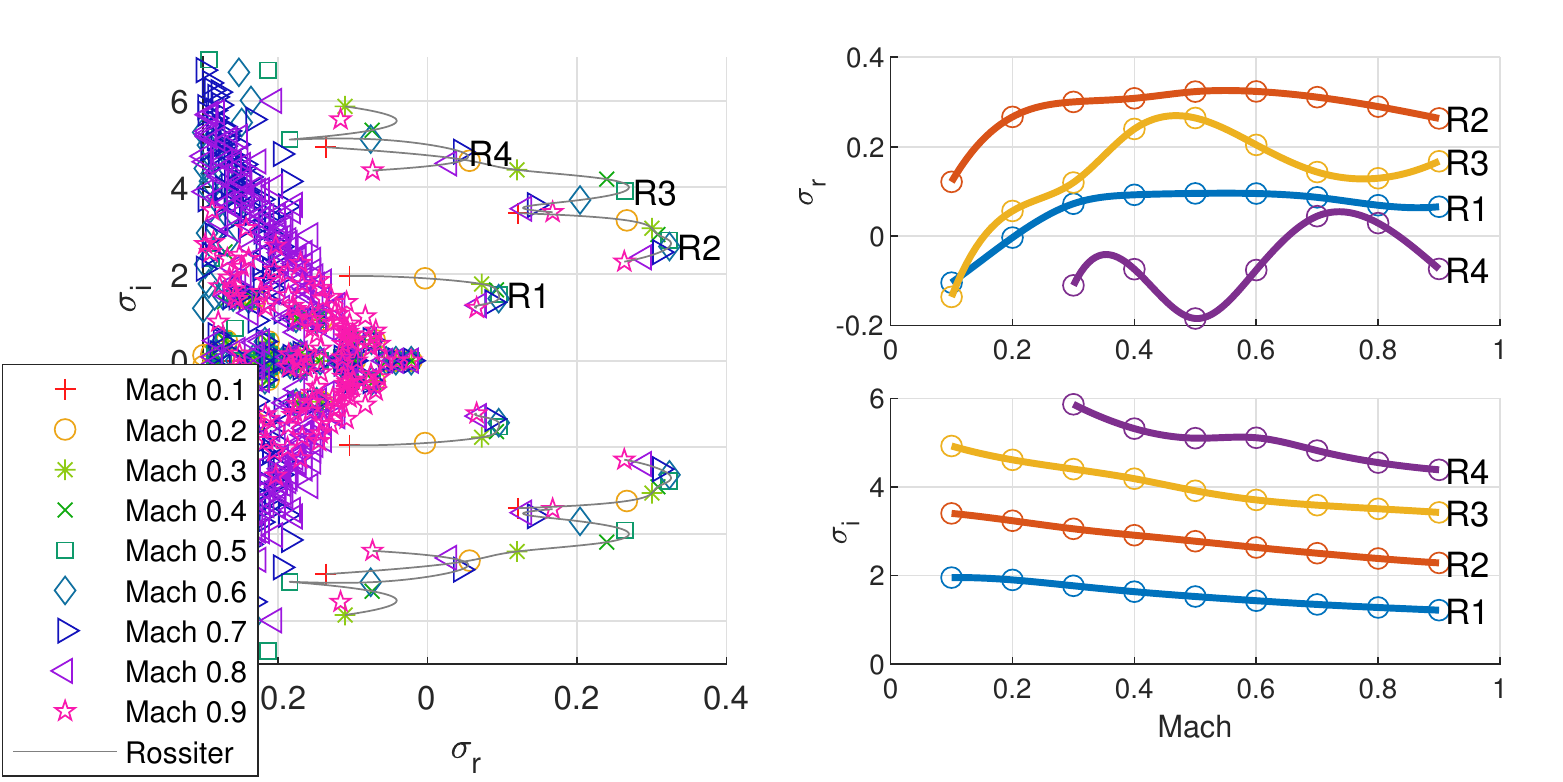}
	\end{center}
	\caption{Eigenvalues from the second Mach number sweep ($Re_D=1000$, $D/\theta=100$). (a) Eigenspectrum. (b) Real and imaginary parts of Rossiter mode eigenvalues.}
	\label{fig:machSweep2}
\end{figure}

For the analysis of the compressibility effects, two representative Mach number sweeps were performed. Figures~\ref{fig:machSweep1}~and~\ref{fig:machSweep2} give, as a function of Mach number, the real and imaginary parts of the eigenvalues for the two $D/\theta$ investigated. The thicker boundary layer at $D/\theta=29.7$ represents a situation where only a pair of Rossiter modes is slightly unstable, while the thinner boundary layer at $D/\theta=100$ is a more complex situation, with up to four unstable modes with higher temporal amplification levels.

Within the parameter range covered, in general the sensitivity of the growth rates with respect to the Mach number was stronger for lower Mach numbers. For low $D/\theta$ (figure~\ref{fig:machSweep1}), at low Mach numbers, both R1 and R2 are stable and only these two modes become unstable as the Mach number increases. At low Mach numbers, R1 is the least stable. It becomes unstable at $Ma=0.3$ and its growth rates saturates at about $Ma=0.4$. R2 is more stable at low Mach numbers, but is sensitive to it, such that it becomes more unstable than R1 for $Ma>0.5$.

For high $D/\theta$ (figure~\ref{fig:machSweep2}), four Rossiter mode linear approximations can be unstable in the Mach number range covered. Mode R2 is the most unstable for all Mach numbers and both R1 and R2 growth rates increase with the Mach number at low numbers, but saturate at about $Ma=0.6$. R3 and R4 are affected by compressibility in a more complex way. In all cases, the frequency reduces with the Mach number, a feature that is predicted by the Rossiter empirical equation and caused by the slower acoustic feedback mechanism.

In summary, the most salient features of the Mach number effect are (1) the strong destabilizing effect of compressibility at low Mach number and (2), at high Mach numbers, either saturation or irregular behavior, depending on mode number and $D/\theta$. 


\subsection{The destabilizing effect of Mach number}

Figures~\ref{fig:machSweep1}~and~\ref{fig:machSweep2} indicate a very strong destabilizing effect of Mach number at low numbers, where compressibility has a very small stabilizing effect on the mixing layer instability. Since the mixing layer profiles were virtually unaffected by the Mach number, the enhancement of the instability could not be associated directly with it. The receptivity of shear layer instability modes is not known to be very sensitive to the Mach number. On the other hand, the transfer of energy into acoustic waves grows very rapidly with the Mach number \cite{Goldstein1976}.

\begin{figure}[h!]
	\begin{center}
		\includegraphics[width=1\textwidth] {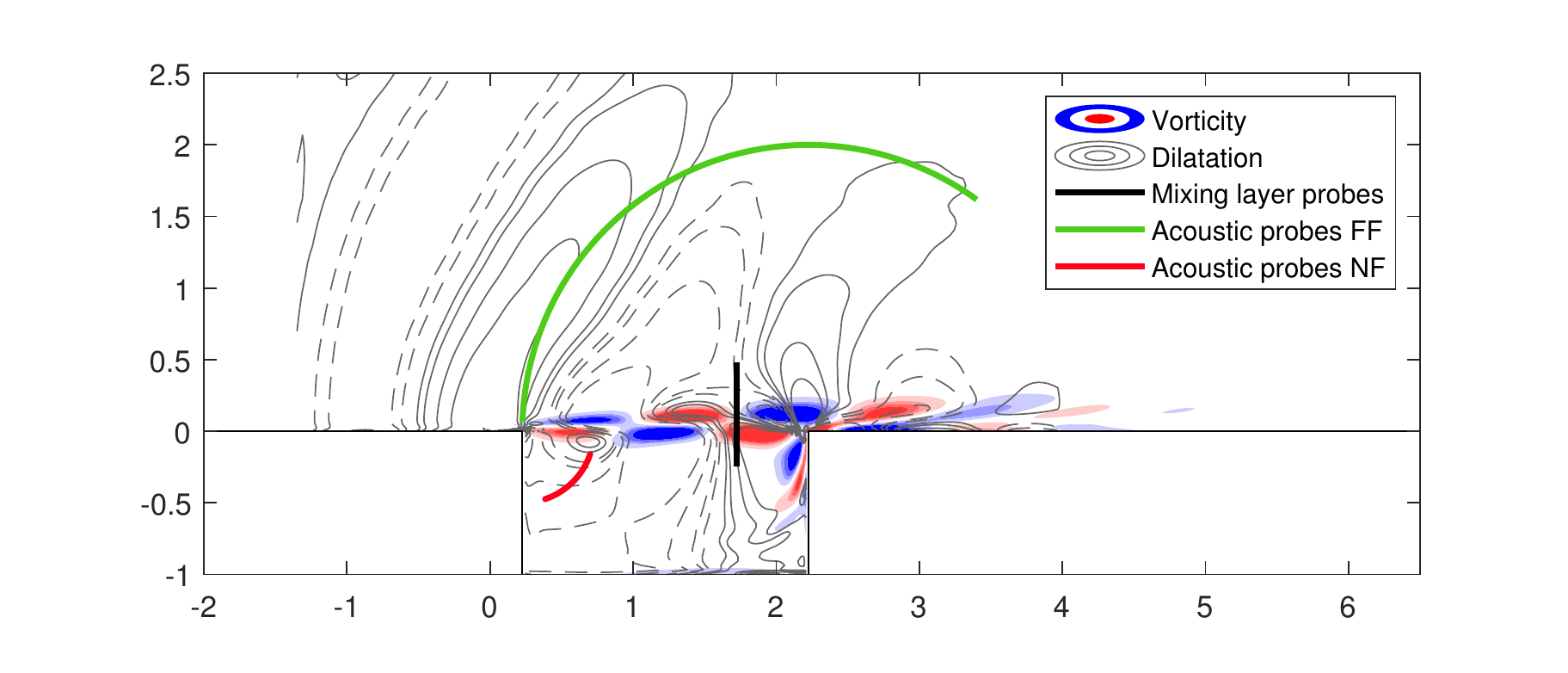}
	\end{center}
	\caption{Location of probes for the analysis of sound emission. Gray contours indicate dilatation fluctuations of the second Rossiter mode at $Ma=0.6$, full line for positive contours and dashed line for negative ones. Colored contours indicate vorticity fluctuations. The black line indicates region of evaluation of sound source magnitude. Red and green lines indicate regions of evaluation of the magnitude of the pressure fluctuations.}
	\label{fig:ReceptivityProbePlacement}
\end{figure}

\citet{Howe2004} investigates the emission of sound by a cavity at low Mach numbers. In Howe's model, the acoustic energy transfer ($ET_{Ac}$) was computed as

\begin{equation}
ET_{Ac}=\frac{E_{Ac}}{E_{SL}} \ ,
\label{eq:acousticReceptivity}
\end{equation}

\noindent where

\begin{equation}
E_{Ac}=\int_{P_{Ac}} \left|p\right|^2 dx \ ,
\label{eq:acousticEnergy}
\end{equation}

\noindent is the acoustic power in the far field and

\begin{equation}
E_{SL}=\int_{P_{SL}} \left| \nabla \cdot \left( \bm{\omega} \wedge \bm{v} \right) \right| dx,
\label{eq:shearLayerEnergy}
\end{equation}

\noindent is the source term. In the equations, $p$, $v$ and $\bm{\omega}$ are respectively pressure, velocity and vorticity. All these quantities can be obtained from the eigenfunctions of the Rossiter mode linear approximations. This was done as illustrated in figure~\ref{fig:ReceptivityProbePlacement}. The source term was estimated by integrating across the shear layers, along the vertical black line shown in the figure. For the acoustic power, the integration was along a semi-circumference centered at the cavity trailing edge with radius $2D$, initiating at the cavity leading edge (the green line). It is not entirely clear whether this region corresponds to the near or the far field of the acoustic source, in particular in view that the extension of these fields is also affected by the Mach number. Regardless of that, we will refer to this region as acoustic field. Clearly, this position provides an estimate of the pressure fluctuations that trigger the Kelvin-Helmholtz vortices. In the picture, isocontours of the values of pressure and velocity eigenfunctions of the Rossiter mode 2 at $Ma=0.6$ were overlaid to illustrate the flow at an arbitrary phase.

\begin{figure}[h!]
	\begin{center}
		\includegraphics[width=0.9\textwidth] {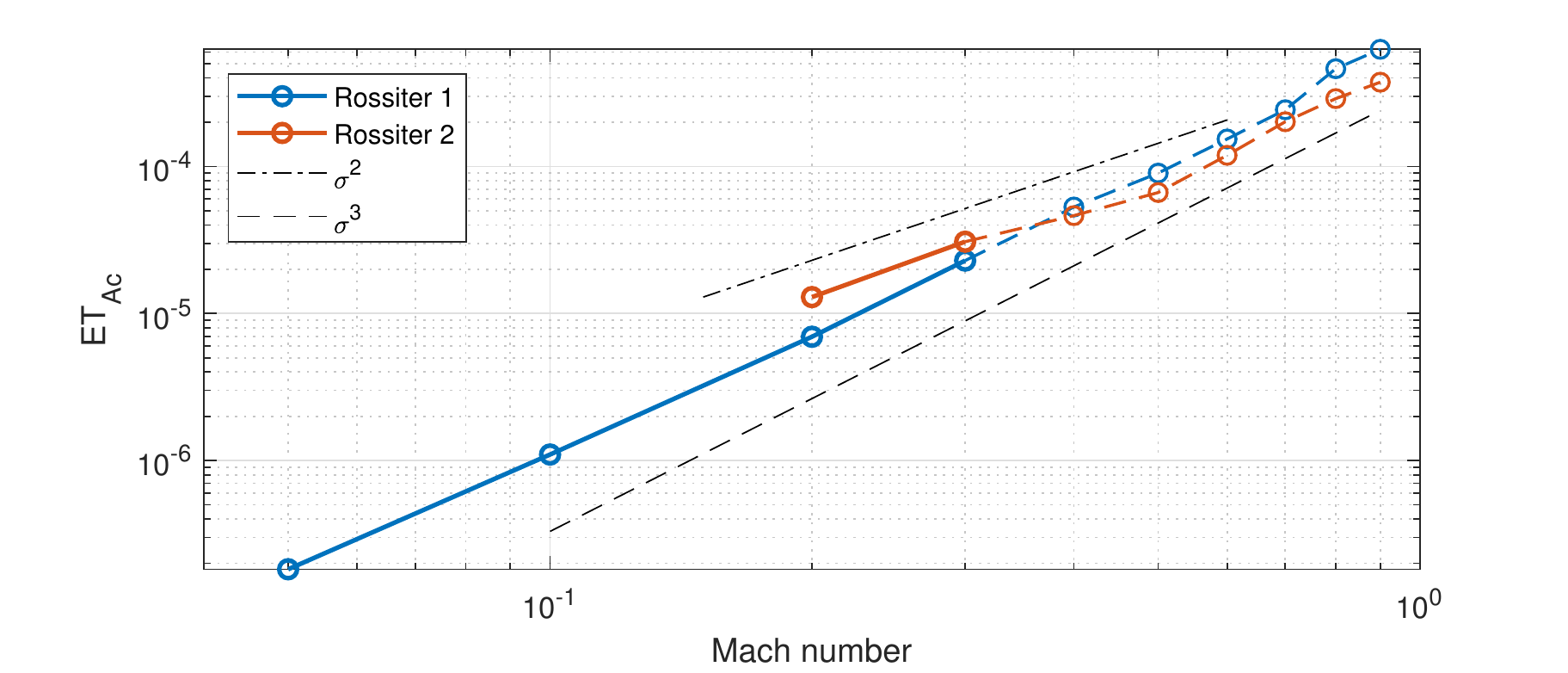}
	\end{center}
	\caption{Acoustic energy transfer ($ET_{AC}$) as a function of Mach number for the linear approximation of Rossiter modes 1 (blue) and 2 (red). The dashed and dashed-dotted lines represent $Ma^3$ and $Ma^2$ respectively.}
	\label{fig:AcousticReceptivity}
\end{figure}

In the analysis, the pressure and the velocities are normalized by their far-field values. Figure~\ref{fig:AcousticReceptivity} shows the $ET_{AC}$ as a function of Mach number, for modes 1 and 2 at $D/\theta=29.7$. The picture includes power functions of exponent 2 and 3 for reference. The acoustic energy transfer increases with the Mach number raised to a power between 2 and 3, depending on the Mach number and the mode.

The evaluation of the source term and acoustic power was also carried out with other integration regions around the green and blue lines of figure~\ref{fig:ReceptivityProbePlacement} and including one inside the cavity around the leading edge to evaluate the acoustic feedback internal to the cavity (marked in red in figure~\ref{fig:ReceptivityProbePlacement}). The Mach number scaling was insensitive to the positions chosen.

The results are consistent with those by \citet{Howe2004}, which suggests that, at most frequencies, a dipole dominates the acoustic emission of the open cavity which is given by $\rho_0 U^3 Ma^3$. However, recall that it is unclear whether our so-called acoustic region corresponds to the near or the far field of the acoustic source. Moreover, Howe's theory is restricted to very low Mach numbers. \cite{Yamouni2013} also presents other arguments for a dipole source in connection with the acoustic cavity modes. In any case, more important for our analysis is the observation that indeed a consistent quantification of energy transfer could be extracted from the eigenfunctions and that the strong dependence on the Mach number offers an explanation for the large sensitivity of the amplification rates of the linear Rossiter mode at low Mach numbers.

\subsection{Peaks and valleys of instability as the Mach number changes}

Figure~\ref{fig:machSweep2} shows that for high $D/\theta$, modes 3 and 4 growth rates depend on Mach number in a very complex way, while modes 1 and 2 display a smooth dependence on Mach for both values of $D/\theta$ considered. \citet{Yamouni2013} has observed a similar complex dependence and linked it to a resonance between Rossiter modes and standing waves in the cavity. These standing waves are described by \citet{Plumblee1962}. 

\begin{figure}[h!]
	\begin{center}
		\includegraphics[width=0.9\textwidth]{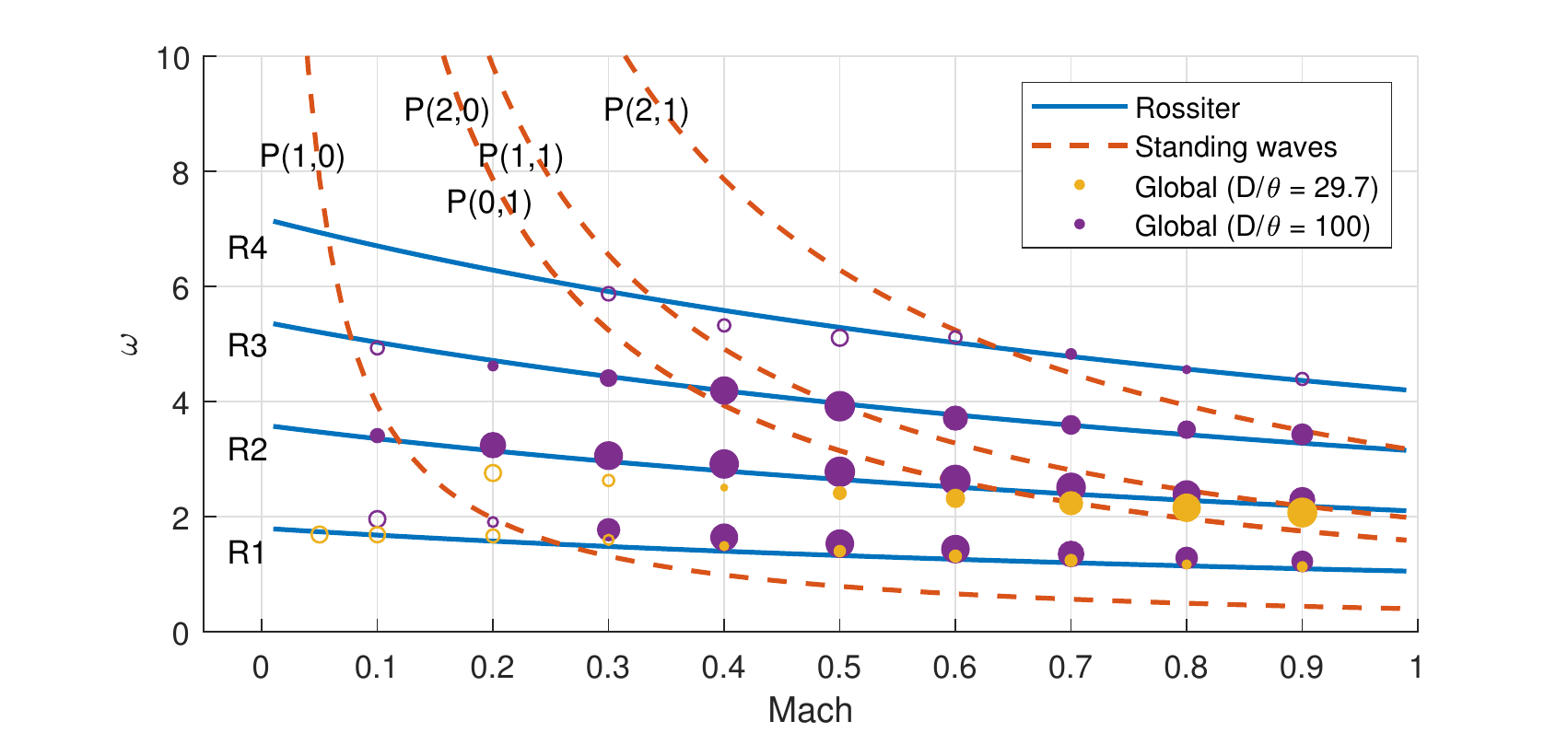}
	\end{center}
	\caption{Frequency predictions of Rossiter modes (blue) and Plumblee acoustic cavity modes (orange) compared to the global mode results (circles). The circle radius is proportional to the distance from neutral stability, the solid circles represent instability while hollow circles represent stability.}
	\label{fig:plumbleeCompare}
\end{figure}

Figure~\ref{fig:plumbleeCompare} presents the frequency of Rossiter and Plumblee modes as a function of Mach number. The blue solid lines represent the Rossiter modes as predicted by 

\begin{equation}
\omega=\frac{2\pi N_R}{\frac{1}{\kappa}+Ma\left(1+\frac{0.514}{L/D}\right)}, 
\label{eq:Rossiter}
\end{equation}

\noindent an equation proposed by \citet{Block1976} and used by \citet{Yamouni2013}, which takes into account the cavity aspect ratio ($L/D$). In the equation, $N_R$ is the Rossiter mode number and $\kappa=1/1.75$ is an empirical constant. In the figure, the dashed orange lines represent the standing wave modes given by \cite{Plumblee1962}

\begin{equation}
\omega=\frac{\pi}{2Ma}\left[\left(\frac{M_p}{L}\right)^2+\left(\frac{N_p}{D}\right)^2\right],
\label{eq:Plumblee}
\end{equation}

\noindent where $M_P$ is the number of standing waves wavelengths in the stream-wise direction and $N_P$, in the wall-normal direction. The modes are identified by $(M_P,N_P)$. In the case of a cavity with aspect ratio $L/D$=2, modes (2,0) and (0,1) coincide in frequency.

The figure also displays results from global instability analysis given by the circles. The circle radius is proportional to the distance from neutral stability conditions, the solid circles represent instability while hollow circles represent stability. Large filled symbols indicate strong instability, large hollow symbols, the opposite. 

A correlation can be found between both types of modes and the global instability results. Rossiter mode 3 shows increased instability when the R3 curve approaches the $P(2,0)$ and $P(1,1)$ curves and, at a higher Mach numbers, the $P(2,1)$ curve. Mode R4 is also affected by these interactions, not only on the growth rates but also in the frequency, the latter may be associated with the fact that this mode is only marginally unstable or stable. Albeit less pronounced, variations in frequency consistent with this argument are also observed for R3.

At lower frequencies, the $P(m,n)$ curves tend to become parallel to the R1 and R2 curves. This smooths out the dependence on Mach number, but may be associated with a maximum instability for R1 at $Ma=0.6$ and low $D/\theta$ (see figure~\ref{fig:machSweep1}), which has also been reported by \citet{Sun2017} at similar conditions. It may also explain why, contrary to all other modes that were observed to saturate at $Ma=0.6$, mode 2 at low $D/\theta$, seems to saturate only at $Ma=0.9$, figure~\ref{fig:machSweep1}.

\section{Nonlinear effects}
\label{sec:nonlinear}

It is important to evaluate to which extent the linear approximation of Rossiter modes represent the Rossiter modes observed if nonlinear terms are considered. To investigate nonlinear effects, we ran 2D simulations for cases selected from both Mach number sweeps. No disturbance was introduced other than the discretization error. Three-dimensional effects could, of course, be important in such nonlinear regimes, but the most salient features of the 2D simulation are likely to be relevant even if three dimensionality were included because the Rossiter modes dominate the flow and are essentially two-dimensional.

\begin{figure}[h!]
 \begin{center}
 \includegraphics[width=\textwidth] {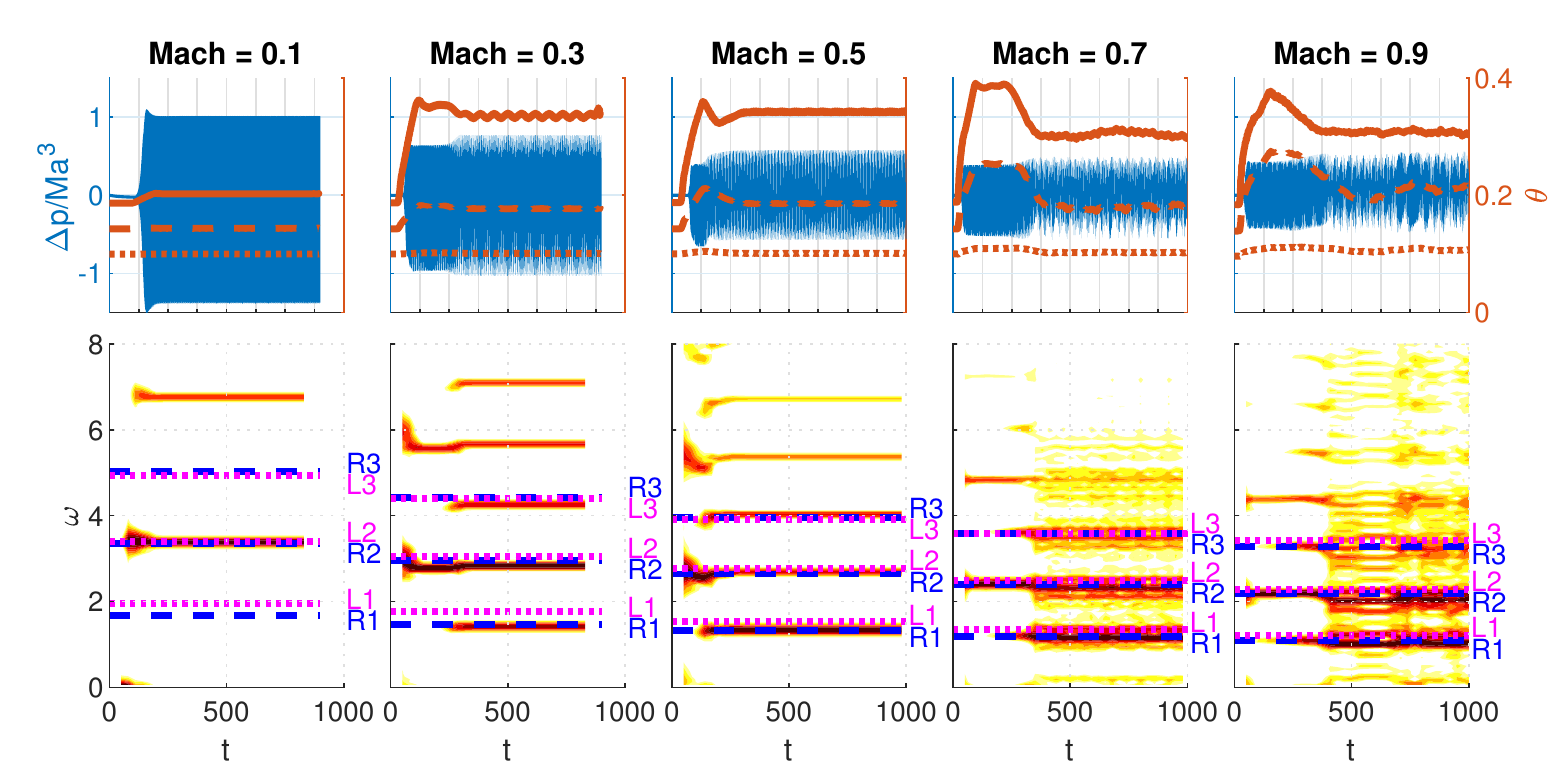}
 \end{center}
 \caption{Two-dimensional DNS results for various Mach numbers at $D/\theta=100$ and $Re_D=1000$. (top) Temporal data of pressure (blue) and mixing layer vorticity thickness (orange). The vorticity thickness data was gathered at positions 1/4 (dotted line), 1/2 (dashed line) and 3/4 (continuous line) of the cavity. (bottom) Dominant frequencies as a function of time for each case. Predictions by the global analysis (dotted magenta, Ln) and by the Rossiter mode empirical equation (dashed blue, Rn) are shown for reference.}
 \label{fig:machFFT100}
\end{figure}

\begin{figure}[h!]
 \begin{center}
 \includegraphics[width=\textwidth] {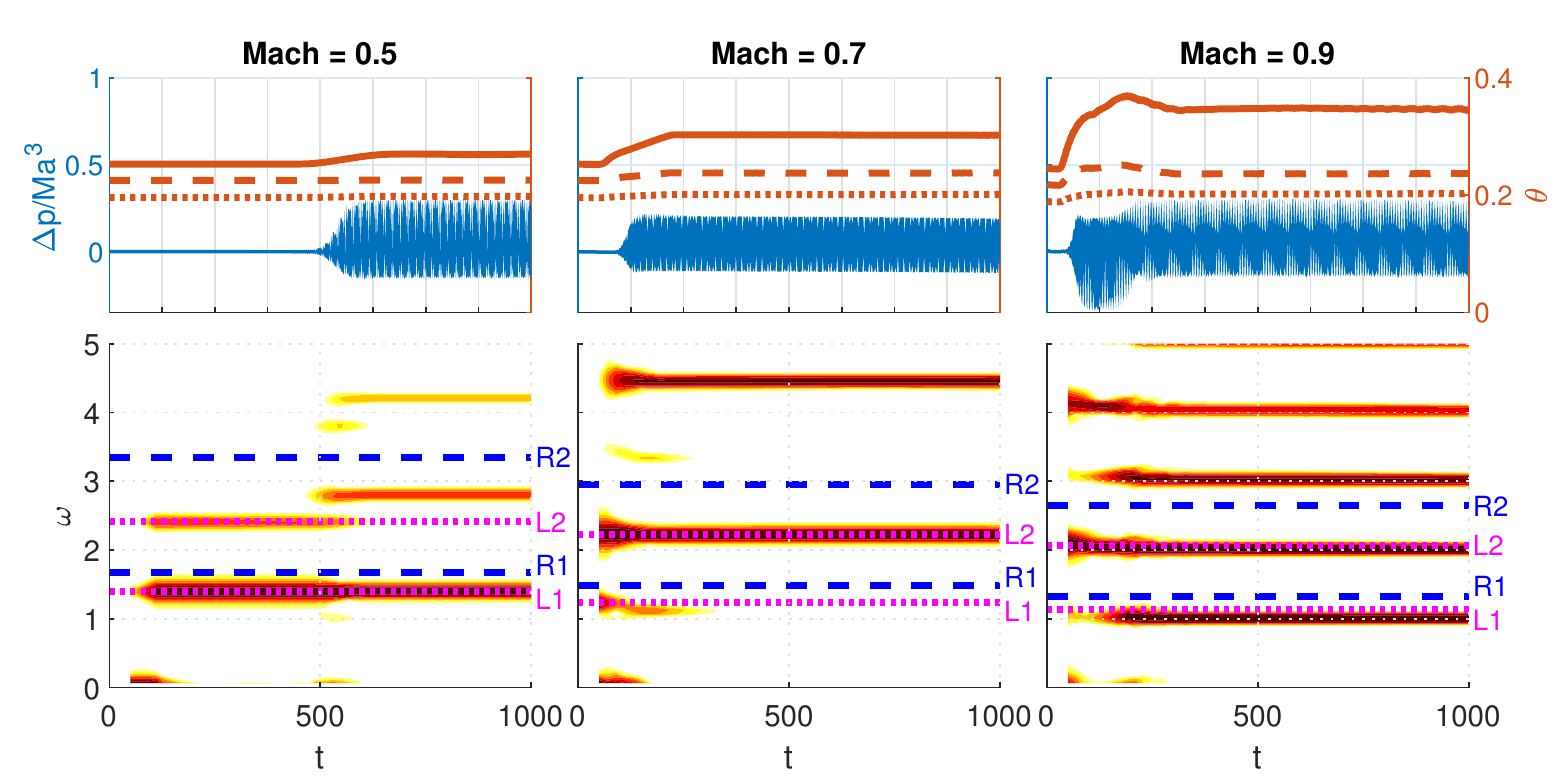}
 \end{center}
 \caption{Two-dimensional DNS results for various Mach numbers at $D/\theta=29.7$ and $Re_D=1149$. (top) Temporal data of pressure (blue) and mixing layer vorticity thickness (orange). The vorticity thickness data was gathered at positions 1/4 (dotted line), 1/2 (dashed line) and 3/4 (continuous line) of the cavity. (bottom) Dominant frequencies as a function of time for each case. Predictions by the global analysis (dotted magenta, Ln) and by the Rossiter mode empirical equation (dashed blue, Rn) are shown for reference.}
 \label{fig:machFFT29}
\end{figure}

Figures~\ref{fig:machFFT100}~and~\ref{fig:machFFT29} display, for both Mach number sweeps, time series of pressure fluctuations at the cavity trailing edge (top frames, blue line) as well as the time evolution of the mixing layer vorticity thickness (top frames, orange lines) at three different stream-wise locations. The bottom frames show, as a function of time, the dominant frequencies contained in the oscillations. The spectra were obtained with the use of moving Hanning window corresponding to 100 simulation time units. The time step in the DNS was $1.37\times10^{-3}$, which was interpolated into a discretization time of $0.1$. The spectra are normalized by the spectral peak for each time window to facilitate visualization. Linear global instability results and empirical frequency predictions of Rossiter modes were added for comparison, respectively indicated by Ln and Rn for the n\textsuperscript{th} Rossiter mode. The empirical predictions are computed by the equation by \citet{Block1976}. The pressure fluctuation in the upper plot is normalized by $Ma^3$ to facilitate the comparison between different Mach number cases and reflect the Mach number effect on the acoustic emission discussed in section~\ref{sec:mach}.

We begin with the thin boundary layer case, figure~\ref{fig:machFFT100} ($D/\theta=100$), because it seems to cover a wider range of regimes. At $Ma=0.1$ the flow closely follows the linear prediction. Initially, only Rossiter mode 2 appears, which was the only one found to be linearly unstable by the bi-global stability analysis. It grows and eventually reaches a limit cycle and generates harmonics, but its frequency matches very well the bi-global (and, for this mode, the empirical) predictions. The mixing layer vorticity thickness remains almost unchanged, except for the last stream-wise position.

$Ma=0.3$ and $Ma=0.5$ with $D/\theta=100$ present results similar to each other. Bi-global analysis predicts modes R1 to R3 to be unstable, mode R2 being the most unstable. In both cases, initially R2 was dominant with a frequency close to the bi-global predictions. Soon after, at about $t=100$, the frequency reduces and approaches the empirical predictions. At the same time, the thickness of the mixing layer increases, which changes the mean velocity profile. At the final stage, mode R1 becomes the dominant. The frequency of this R1 mode is close to the empirical predictions. More R1 is accompanied by its harmonics and $Ma=0.5$ reaches the final stage sooner than $Ma=0.3$

For $Ma=0.7$ and $Ma=0.9$ with $D/\theta=100$, bi-global analysis predicts R1 to R3 to be unstable and, for $Ma=0.7$, R4 is also unstable. For both cases, R2 is predicted as the most unstable. In the nonlinear simulations, only R2 is seen initially and displays a time interval with limit cycle oscillation. Eventually it exhibits a more irregular behavior, with modes R1 and R3 setting in. For these $Ma$, modes R1 and R3 take long to appear in comparison with $Ma=0.5$, but behave in a more complicated way. The final stages for $Ma=0.7$ and $Ma=0.9$ are the most irregular observed. The $Ma=0.7$ case evolves more quickly into the more irregular regime. It is unclear whether this could be associated with the fact that this Mach number has 4 unstable modes while $Ma=0.9$ has only 3. The irregular regime has a more distributed spectra, but there are dominant modes that match the empirical predictions. Once more, the mixing layer thickness and the velocity profiles change in time and the onset of irregular behavior was associated with a change in mean profile. The limit cycle oscillation time interval corresponded to the largest modification. In the final irregular stage, both $Ma=0.7$ and $0.9$ settle to an intermediary level of distortion.

For the thicker boundary layer ($D/\theta=29.7$), shown in figure~\ref{fig:machFFT29}, the linear stability theory predicts that only modes R1 and R2 are unstable. The first important observation is that, for all Mach numbers, the empirical predictions do not agree with the linear stability results. This is likely because the empirical model was based on thin boundary layer experiments. For $Ma<0.3$ the flow is stable. At $Ma=0.5$, initially the oscillations are small and linear and the frequencies consistently match the linear stability predictions. The linear analysis predicts modes R1 and R2 with very similar growth rates and indeed the amplitude ratio between the modes remain constant. It is unclear why mode R1 reaches a larger amplitude than R2, but it may be associated with the uncontrolled initial disturbance. As the amplitude grows a mean flow distortion arises at position 3/4 of the cavity length ($t\approx500$). At this point mode R1 reduces amplitude and  R2 vanishes, while the harmonics of R1 rise. The thicker mixing layer is expected to favor lower Rossiter modes, which may explain the final stage. 

At $Ma=0.7$ mode R2 is the most unstable, and also more unstable that at $Ma=0.5$. At the very beginning of the simulation, both R1 and R2 modes are visible in the spectrum and match the linear predictions. Consistently with being more unstable than $Ma=0.3$, this case develops faster and presents a greater change in the mean flow. As the mean flow changes, only R2 remains, also displaying harmonics in the spectrum. The fact that R2 dominates the flow agrees with the linear prediction as well.

The most unstable case of the thicker boundary layer, $Ma=0.9$, has a more complex behavior and more variation of the mean flow. Initially, mode R2 dominates, in accordance with theory, but soon after, mode R1 appears. However, its frequency does not match perfectly the theory, it is better described as the subharmonic of mode R2. This observation poses a question about the effective origin of mode R1 for the thick boundary layer at high Mach. The final complex stage display at least 4 modes in harmonic order.

Clearly, the nonlinear regime of these instabilities is very complicated, but the are some patterns. For all cases, the linear theory provides good predictions for the initial stages of the mode evolution, both in frequency and dominant mode. As nonlinearity sets is, a limit cycle is formed. This stage, in general, is dominated by the most unstable flow as predicted by the theory. For the thick boundary layer case, the frequencies in this regime are well predicted by the theory, while, for the thin boundary layer scenario, they agree better with the empirical model. In this regime, there is substantial mean flow distortion, which may be associated to the departure from linear theory. After the limit cycle, another regime can occur, which is more complex and irregular. The frequencies remain the same of the limit cycle stage, but, despite not being the most linearly unstable mode, the R1 mode tends to dominate and generate harmonics. This later regime is associated with a reduction in the mean flow distortion. In summary, the frequencies of the complex final stages of the flow agree with the Rossiter empirical predictions and their spectra seem to represent the nonlinearly generated harmonics of the R1 mode, with limited reminiscence of the linear instability that triggered this unsteady flow.

The origin of the R1 mode is unclear, but two possibilities exist: As the flow evolves nonlinearly, a significant thickening of the mixing layer occurs, which, from the analysis of the effect of $D/\theta$ (see also \cite{Mathias2018b}), would favor lower order Rossiter modes. Another possibility are vortex pairings of the mode R2. These mechanism are not mutually exclusive and both take place in a spatial evolution of mixing layers.

The global instability analysis indicates the flow becomes more unstable as Mach number increases up to $Ma=0.5$. Above that, it is unclear that the flow becomes even more unstable because the growth rates either saturate or oscillate with the Mach number, in particular for the thin boundary layer. Nonetheless, it can be said that, in most instances, as the flow becomes more unstable the flow dynamics becomes increasingly more complicated and that the final nonlinear stage is reached more quickly. These features are generally consistent with a weakly nonlinear process \cite{Drazin2004}. 

For all cases the nonlinear simulations indicate progressively larger pressure oscillations as the Mach number increases, which is consistent with the analysis in section~\ref{sec:mach}. However, the limit-cycle amplitude increases with less than $Ma^3$, suggesting other effects of Mach number are present.

\section{Final remarks}
\label{sec:conclusion}

The open cavity flow is governed by several parameters. Large values of $L/D$ promote the wake mode. Since our focus was the Rossiter mode, we fixed $L/D$ at 2. A study of the literature revealed that the effect of $D/\theta$ has been overlooked. Only \citet{Yamouni2013} present results of a sweep of this parameter. However, this was not the main focus of that work, hence the analysis was rather superficial, and their only conclusion was that the flow stabilized as $D/\theta$ reduced. Moreover, they focus on very high Reynolds number, far from the critical conditions. Here, we performed a more in depth analysis and focus on conditions close to critical where the biglobal linear stability analysis is expected to be more meaningful. Our analysis confirmed the argument that the flow stabilized as $D/\theta$ reduces, as expected from the mixing layer instability. However, we went further to show that the Rossiter mode selection and their hierarchy (order of dominance) is essentially governed by the instability of the mixing layer. This was demonstrated by comparison of the global instability results with results from spatial linear stability of the mixing layer. Accordingly, at low $D/\theta$, modes R1 and R2 compete for dominance, while at large $D/\theta$, mode R2 dominates, followed closely by R3, with R1 also unstable, but far behind.

Since two scenarios were established (low and high $D/\theta$), we chose a representative $D/\theta$ for each and performed a Mach number sweep from 0.1 and 0.9. Mach number sweeps have been presented previously for both large and low $D/\theta$, but not for the same $L/D$ and Reynolds number, which blurs any analysis of the effect of $D/\theta$. Mach number sweeps close to critical conditions were carried out for low $D/\theta$, while \cite{Yamouni2013} have carried out a Mach sweep for $D/\theta=231$ and at a large Reynolds  number, far from the critical conditions. They concluded that the effect of Mach number resulted from a competition between the stabilizing effect of compressibility on the mixing layer and the destabilizing effect of resonances between the Rossiter modes and acoustic cavity modes, producing a very complex and irregular dependence on Mach number. We found a different picture. In our parameter space, Mach was massively destabilizing for all modes at low Mach number reaching a saturation at about $Ma=0.5$. Only at large Mach numbers, higher modes displayed an irregular dependence, similar to that observed by \citet{Yamouni2013}, but less intense and restricted to higher order modes. By analyzing the eigenfunctions of the Rossiter modes, we established the amount of energy that is transferred from the vorticity field to the acoustic field of the mode. We obtained that this energy transfer increases approximately with $Ma^3$, in agreement with simplified models of cavity noise emission that apply to this scenario \cite{Howe2004}. This explained the strong destabilizing effect of Mach number at the low range. The power law dictates that as Mach number increases the effect of an identical increment must reduce. This is the main reason for the observed reduction of the effect of Mach number as it increases. On the other hand, the stabilizing effect on the mixing layer of increasing the Mach number in the subsonic regime is also likely to contribute.

The irregular dependence on Mach number at high subsonic values was traced to resonances with the acoustic modes, showing that the phenomenon described by \citet{Yamouni2013} at large $Re$ is also active close to critical conditions. However, some differences were observed. For high Rossiter modes (R3 and R4) the lines governing the Rossiter and the acoustic modes on a $Ma \times \omega$ plane cross each other at well defined points, indicating distinct resonances which affect the instability. For lower order Rossiter (R1 and R2), these lines tend to become parallel and the points of resonance become ill defined. For mode R2, the resonance is active over a wide range of Mach numbers. As a consequence, for $D/\theta=29.7$, mode R2 becomes progressively more unstable as $Ma$ increases up to 0.9 as opposed to mode R1 which is little affected by such resonances and saturates at $Ma=0.5$. Therefore, for this $D/\theta$, at about $Ma=0.6$, mode 2 becomes dominant. \citet{Sun2016} investigated the effect of Mach number on the instability of a cavity at $Re_D=1500$, $L/D=2$ and $D/\theta=26.5$, parameters very similar to our low $D/\theta$ case. They also observed that at $Ma\approx0.6$, the dominant mode switches from R1 to R2, a feature they could not explain. In view of the great similarity with our parameters, this is almost certainly associated with the resonance effects discussed. With these analyses, we were able establish the physical mechanisms that govern the instability and explain the mode selection and hierarchy throughout the parameter space covered.

Having performed the linear stability, we then verified to which extend the linear results can predict the nonlinear saturated limit of the Rossiter instability. For that purpose, we performed numerical simulations. Comparison of linear stability results and nonlinear simulation results were reported by \citet{Sun2016}, both in 2D and in 3D, but only for one set of flow parameters. We performed DNS simulations for the whole range of $D/\theta$ covered in the linear analysis. They represent regions with different hierarchy and number of Rossiter modes as well as different levels of instability. The simulations were 2D, but as discussed in the paper, they are expected to display the most salient nonlinear features. Analysis of the nonlinear instability indicated a very complex flow. Initially, the flow behaved accordingly to linear theory, but, in the nonlinear regime, the behavior was progressively more complex for the more unstable cases. The final stage tended to be dominated by the R1 mode, which was not the most linearly unstable, and its harmonics. At this stage, for thin boundary layers, the empirical model provided better predictions of mode frequency than the linear theory. The origin of the R1 mode is unclear, but two possibilities exist. As the flow evolves nonlinearly, a significant thickening of the mixing layer occurs, which, from the analysis of the effect of $D/\theta$ would favor lower order Rossiter modes (see also \citet{Mathias2018b}). Another possibility is the vortex pairing of the mode R2. These mechanisms are not mutually exclusive, and both take place in a spatial evolution of mixing layers.

In summary we selected an $L/D$ which was representative of the scenario where Rossiter modes dominate. For this parameter we performed a $D/\theta$ sweep and established two scenarios, namely, low and high $D/\theta$. Finally, we performed a Mach number sweep for a representative $D/\theta$ of each scenario. Our analysis focus on Reynolds numbers close to critical conditions. However, \citet{Yamouni2013} investigated one of our $D/\theta$ cases at a very high Reynolds number and found 6 unstable modes, rather than 4, but no additional physics took place. In view of this, it can be said that our study provides a first comprehensive analysis of the effects of both Mach number and the ratio $D/\theta$ on the two dimensional linear and nonlinear instability of Rossiter modes in subsonic flows.

\section*{Acknowledgements}
	
The authors would like to thank the São Paulo Research Foundation (FAPESP/Brazil), for grants 2018/04584-0 and 2017/23622-8; the National Council for Scientific and Technological Development (CNPq/Brazil) for grants 134722/2016-7 and 307956/2019-9; the US Air Force Office of Scientific Research (AFOSR) for grant FA9550-18-1-0112, managed by Dr. Geoff Andersen from SOARD; the University of Liverpool for the access to the Barkla cluster, provided by Prof. Vassilios Theofilis; and the Center for Mathematical Sciences Applied to Industry (CeMEAI) funded by São Paulo Research Foundation (FAPESP/Brazil), grant \#2013/07375-0, for access to the Euler cluster, provided by Prof. José Alberto Cuminato.

%
%
\bibliographystyle{plainnat} 
\bibliography{library} 

\begin{thebibliography}{44}
\providecommand{\natexlab}[1]{#1}
\providecommand{\url}[1]{\texttt{#1}}
\expandafter\ifx\csname urlstyle\endcsname\relax
  \providecommand{\doi}[1]{doi: #1}\else
  \providecommand{\doi}{doi: \begingroup \urlstyle{rm}\Url}\fi

\bibitem[{\AA}kervik et~al.(2006){\AA}kervik, Brandt, Henningson, H{\oe}pffner,
  Marxen, and Schlatter]{Akervik2006}
Espen {\AA}kervik, Luca Brandt, Dan~S. Henningson, J{\'{e}}rome H{\oe}pffner,
  Olaf Marxen, and Philipp Schlatter.
\newblock {Steady solutions of the Navier-Stokes equations by selective
  frequency damping}.
\newblock \emph{Physics of Fluids}, 18\penalty0 (6):\penalty0 68--102, 2006.
\newblock ISSN 10706631.
\newblock \doi{10.1063/1.2211705}.
\newblock URL
  \url{http://scitation.aip.org/content/aip/journal/pof2/18/6/10.1063/1.2211705}.

\bibitem[Arnoldi(1951)]{Arnoldi1951}
W.~E. Arnoldi.
\newblock {The principle of minimized iterations in the solution of the matrix
  eigenvalue problem}.
\newblock \emph{Quarterly of Applied Mathematics}, 9\penalty0 (1):\penalty0
  17--29, 1951.

\bibitem[Bergamo et~al.(2015)Bergamo, Gennaro, Theofilis, and
  Medeiros]{Bergamo2015}
Leandro~F. Bergamo, Elmer~M. Gennaro, Vassilis Theofilis, and Marcello~A.F.
  Medeiros.
\newblock {Compressible modes in a square lid-driven cavity}.
\newblock \emph{Aerospace Science and Technology}, 44:\penalty0 125--134, jul
  2015.
\newblock ISSN 12709638.
\newblock \doi{10.1016/j.ast.2015.03.010}.
\newblock URL
  \url{https://linkinghub.elsevier.com/retrieve/pii/S1270963815001030}.

\bibitem[Block(1976)]{Block1976}
Patricia J.~W. Block.
\newblock {Noise response of cavities of varying dimensions at subsonic
  speeds}.
\newblock Technical Report Nasa Techinical Note D-8351, National Aeronautics
  and Space Administration, 1976.

\bibitem[Br{\`{e}}s and Colonius(2008)]{Bres2008}
Guillaume~A. Br{\`{e}}s and Tim Colonius.
\newblock {Three-dimensional instabilities in compressible flow over open
  cavities}.
\newblock \emph{Journal of Fluid Mechanics}, 599:\penalty0 309--339, mar 2008.
\newblock ISSN 0022-1120.
\newblock \doi{10.1017/S0022112007009925}.
\newblock URL
  \url{http://www.journals.cambridge.org/abstract{\_}S0022112007009925}.

\bibitem[Chiba(1998)]{Chiba1998}
S.~Chiba.
\newblock {Global Stability Analysis of Incompressible Viscous Flow}.
\newblock \emph{Journal of Japan Society of Computational Fluid Dynamics},
  7\penalty0 (1):\penalty0 20--48, 1998.

\bibitem[Citro et~al.(2015)Citro, Giannetti, Brandt, and Luchini]{Citro2015}
Vincenzo Citro, Flavio Giannetti, Luca Brandt, and Paolo Luchini.
\newblock {Linear three-dimensional global and asymptotic stability analysis of
  incompressible open cavity flow}.
\newblock \emph{Journal of Fluid Mechanics}, 768:\penalty0 113--140, 2015.
\newblock ISSN 0022-1120.
\newblock \doi{10.1017/jfm.2015.72}.
\newblock URL
  \url{http://www.journals.cambridge.org/abstract{\_}S0022112015000725}.

\bibitem[Colonius et~al.(1999)Colonius, Basu, and Rowley]{Colonius1999}
Tim Colonius, Amit~J. Basu, and Clarence~W. Rowley.
\newblock {Computation of sound generation and flow-acoustic instabilities in
  the flow past an open cavity}.
\newblock In \emph{Proceedings of the Joint Fluids Engineering Conference}, San
  Francisco, USA, 1999.

\bibitem[de~Vicente et~al.(2014)de~Vicente, Basley, Meseguer-Garrido, Soria,
  and Theofilis]{DeVicente2014}
J.~de~Vicente, J.~Basley, F.~Meseguer-Garrido, Julio Soria, and Vassilios
  Theofilis.
\newblock {Three-dimensional instabilities over a rectangular open cavity: from
  linear stability analysis to experimentation}.
\newblock \emph{Journal of Fluid Mechanics}, 748:\penalty0 189--220, 2014.
\newblock ISSN 0022-1120.
\newblock \doi{10.1017/jfm.2014.126}.
\newblock URL
  \url{http://www.journals.cambridge.org/abstract{\_}S0022112014001268}.

\bibitem[Dix and Bauer(2000)]{Dix2000b}
R.~E. Dix and R.~C. Bauer.
\newblock {Experimental and Theoretical Study of Cavity Acoustics}.
\newblock Technical report, Sverdrup Technology, Inc./ AEDC Group for Arnold
  Air Force Base, Tennessee, USAF, 2000.

\bibitem[Drazin and Reid(2004)]{Drazin2004}
P.~G. Drazin and W.~H. Reid.
\newblock \emph{{Hydrodynamic Stability}}.
\newblock Cambridge University Press, aug 2004.
\newblock ISBN 9780521525411.
\newblock \doi{10.1017/CBO9780511616938}.
\newblock URL
  \url{https://www.cambridge.org/core/product/identifier/9780511616938/type/book}.

\bibitem[East(1966)]{East1966}
L.F. East.
\newblock {Aerodynamically induced resonance in rectangular cavities}.
\newblock \emph{Journal of Sound and Vibration}, 3\penalty0 (3):\penalty0
  277--287, 1966.
\newblock ISSN 0022460X.
\newblock \doi{10.1016/0022-460X(66)90096-4}.

\bibitem[Gaitonde and Visbal(1998)]{Gaitonde1998}
Datta~V. Gaitonde and Miguel~R. Visbal.
\newblock {High-Order Schemes for Navier-Stokes Equations: Algorithm and
  Implementation Into FDL3DI}.
\newblock Technical report, Wright-Patterson Air Force Base, 1998.

\bibitem[Germanos et~al.(2009)Germanos, {De Souza}, and {De
  Medeiros}]{Germanos2009}
Ricardo~A.Coppola Germanos, Leandro~Franco {De Souza}, and Marcello~A.Faraco
  {De Medeiros}.
\newblock {Numerical investigation of the three-dimensional secondary
  instabilities in the time-developing compressible mixing layer}.
\newblock \emph{Journal of the Brazilian Society of Mechanical Sciences and
  Engineering}, 31\penalty0 (2):\penalty0 125--136, 2009.
\newblock ISSN 18063691.
\newblock \doi{10.1590/S1678-58782009000200005}.

\bibitem[Gharib and Roshko(1987)]{Gharib1987}
M.~Gharib and Anatol Roshko.
\newblock {The effect of flow oscillations on cavity drag}.
\newblock \emph{Journal of Fluid Mechanics}, 177:\penalty0 501, apr 1987.
\newblock ISSN 0022-1120.
\newblock \doi{10.1017/S002211208700106X}.
\newblock URL
  \url{http://www.journals.cambridge.org/abstract{\_}S002211208700106X}.

\bibitem[Goldstein(1976)]{Goldstein1976}
M~E Goldstein.
\newblock \emph{{Aeroacoustics}}.
\newblock McGraw-Hill International Book Company, 1976.
\newblock ISBN 9780070236851.

\bibitem[G{\'{o}}mez et~al.(2015)G{\'{o}}mez, P{\'{e}}rez, Blackburn, and
  Theofilis]{Gomez2015a}
Francisco G{\'{o}}mez, Jos{\'{e}}~Miguel P{\'{e}}rez, Hugh~M. Blackburn, and
  Vassilios Theofilis.
\newblock {On the use of matrix-free shift-invert strategies for global flow
  instability analysis}.
\newblock \emph{Aerospace Science and Technology}, 44:\penalty0 69--76, jul
  2015.
\newblock ISSN 12709638.
\newblock \doi{10.1016/j.ast.2014.11.003}.
\newblock URL
  \url{http://linkinghub.elsevier.com/retrieve/pii/S1270963814002284}.

\bibitem[Howe(2004)]{Howe2004}
M.~S. Howe.
\newblock {Mechanism of sound generation by low Mach number flow over a wall
  cavity}.
\newblock \emph{Journal of Sound and Vibration}, 273\penalty0 (1-2):\penalty0
  103--123, 2004.
\newblock ISSN 0022460X.
\newblock \doi{10.1016/S0022-460X(03)00644-8}.
\newblock URL
  \url{http://linkinghub.elsevier.com/retrieve/pii/S0022460X03006448}.

\bibitem[Juniper et~al.(2014)Juniper, Hanifi, and Theofilis]{Juniper2014}
Matthew~P. Juniper, Ardeshir Hanifi, and Vassilios Theofilis.
\newblock {Modal Stability Theory Lecture notes from the FLOW-NORDITA Summer
  School on Advanced Instability Methods for Complex Flows, Stockholm, Sweden,
  2013 1}.
\newblock \emph{Applied Mechanics Reviews}, 66\penalty0 (2):\penalty0 021004,
  mar 2014.
\newblock ISSN 0003-6900.
\newblock \doi{10.1115/1.4026604}.
\newblock URL
  \url{http://appliedmechanicsreviews.asmedigitalcollection.asme.org/article.aspx?doi=10.1115/1.4026604}.

\bibitem[Krishnamurty(1956)]{Krishnamurty1956}
K.~Krishnamurty.
\newblock {Acoustic Radiation from Two-dimensional Rectangular Cutouts in
  Aerodynamic Surfaces}.
\newblock Technical Report Naca Technical Note - NACA-TN(3487):34, National
  Advisory Committee for Aeronautics, Washington, 1956.

\bibitem[Lele(1992)]{Lele1992}
Sanjiva~K. Lele.
\newblock {Compact finite difference schemes with spectral-like resolution}.
\newblock \emph{Journal of Computational Physics}, 103\penalty0 (1):\penalty0
  16--42, 1992.
\newblock ISSN 00219991.
\newblock \doi{10.1016/0021-9991(92)90324-R}.

\bibitem[Li and Laizet(2010)]{Li2010}
Ning Li and Sylvain Laizet.
\newblock {2DECOMP and FFT-A Highly Scalable 2D Decomposition Library and FFT
  Interface}.
\newblock \emph{Cray User Group 2010 conference}, pages 1--13, 2010.

\bibitem[Martinez and Medeiros(2016)]{Martinez2016}
Andres Martinez and Marcello~F. Medeiros.
\newblock {Direct numerical simulation of a wavepacket in a boundary layer at
  Mach 0.9}.
\newblock In \emph{46th AIAA Fluid Dynamics Conference}, volume 414, pages
  1--33, Reston, Virginia, jun 2016. American Institute of Aeronautics and
  Astronautics.
\newblock ISBN 978-1-62410-436-7.
\newblock \doi{10.2514/6.2016-3195}.
\newblock URL \url{http://arc.aiaa.org/doi/10.2514/6.2016-3195}.

\bibitem[Mathias and Medeiros(2018{\natexlab{a}})]{Mathias2018b}
Marlon Mathias and Marcello~F. Medeiros.
\newblock {The Influence of the Boundary Layer Thickness on the Stability of
  the Rossiter Modes of a Compressible Rectangular Cavity}.
\newblock In \emph{2018 Fluid Dynamics Conference}, Reston, Virginia, jun
  2018{\natexlab{a}}. American Institute of Aeronautics and Astronautics.
\newblock ISBN 978-1-62410-553-1.
\newblock \doi{10.2514/6.2018-3386}.
\newblock URL \url{https://arc.aiaa.org/doi/10.2514/6.2018-3386}.

\bibitem[Mathias and Medeiros(2018{\natexlab{b}})]{Mathias2018}
Marlon~Sproesser Mathias and Marcello Medeiros.
\newblock {Direct Numerical Simulation of a Compressible Flow and Matrix-Free
  Analysis of its Instabilities over an Open Cavity}.
\newblock \emph{Journal of Aerospace Technology and Management}, 10:\penalty0
  1--13, jul 2018{\natexlab{b}}.
\newblock ISSN 2175-9146.
\newblock \doi{10.5028/jatm.v10.949}.
\newblock URL \url{http://www.jatm.com.br/ojs/index.php/jatm/article/view/949}.

\bibitem[McGregor and White(1970)]{Mcgregor1970}
O.~W. McGregor and R.A. White.
\newblock {Drag of rectangular cavities in supersonic and transonic flow
  including the effects of cavity resonance}.
\newblock \emph{AIAA Journal}, 8\penalty0 (11):\penalty0 1959--1964, nov 1970.
\newblock ISSN 0001-1452.
\newblock \doi{10.2514/3.6032}.
\newblock URL \url{http://arc.aiaa.org/doi/abs/10.2514/3.6032}.

\bibitem[Meseguer-Garrido et~al.(2014)Meseguer-Garrido, de~Vicente, Valero, and
  Theofilis]{Meseguer2014}
F.~Meseguer-Garrido, J.~de~Vicente, E.~Valero, and Vassilios Theofilis.
\newblock {On linear instability mechanisms in incompressible open cavity
  flow}.
\newblock \emph{Journal of Fluid Mechanics}, 752:\penalty0 219--236, 2014.
\newblock ISSN 0022-1120.
\newblock \doi{10.1017/jfm.2014.253}.
\newblock URL
  \url{http://journals.cambridge.org/abstract{\_}S0022112014002535}.

\bibitem[Miles(1958)]{Miles1958}
John~W. Miles.
\newblock {On the disturbed motion of a plane vortex sheet}.
\newblock \emph{Journal of Fluid Mechanics}, 4\penalty0 (5):\penalty0 538--552,
  1958.
\newblock ISSN 14697645.
\newblock \doi{10.1017/S0022112058000653}.

\bibitem[Ohmichi and Suzuki(2016)]{Ohmichi2016}
Y.~Ohmichi and K.~Suzuki.
\newblock {Assessment of global linear stability analysis using a time-stepping
  approach for compressible flows}.
\newblock \emph{International Journal for Numerical Methods in Fluids},
  80\penalty0 (10):\penalty0 614--627, apr 2016.
\newblock ISSN 02712091.
\newblock \doi{10.1002/fld.4166}.
\newblock URL \url{http://doi.wiley.com/10.1002/fld.4166}.

\bibitem[Owen(1958)]{Owen1958}
T.~B. Owen.
\newblock {Techniques of pressure fluctuation measurements}.
\newblock Technical Report Advisory Group for Aeronautical Research and
  Development Report 172, North Atlantic Treaty Organization, 1958.

\bibitem[Pavithran and Redekopp(1989)]{Pavithran1989}
S.~Pavithran and L.~G. Redekopp.
\newblock {The absolute-convective transition in subsonic mixing layers}.
\newblock \emph{Physics of Fluids A}, 1\penalty0 (10):\penalty0 1736--1739,
  1989.
\newblock ISSN 08998213.
\newblock \doi{10.1063/1.857496}.

\bibitem[Plumblee et~al.(1962)Plumblee, Gibson, and Lassiter]{Plumblee1962}
H.~E. Plumblee, J.~S. Gibson, and L.~W. Lassiter.
\newblock {Theoretical and Experimental Investigation of The Acoustic Response
  of Cavities In An Aerodynamic Flow}.
\newblock Technical Report USAF Report WADD-TR-61-75, Wright-Patterson Air
  Force Base, 1962.

\bibitem[Ragab and Wu(1989)]{Ragab1989}
Saad~A. Ragab and J.~L. Wu.
\newblock {Linear instabilities in two-dimensional compressible mixing layers}.
\newblock \emph{Physics of Fluids A}, 1\penalty0 (6):\penalty0 957--966, 1989.
\newblock ISSN 08998213.
\newblock \doi{10.1063/1.857407}.

\bibitem[Rossiter(1964)]{Rossiter1964}
J.~E. Rossiter.
\newblock {Wind-tunnel experiments on the flow over rectangular cavities at
  subsonic and transonic speeds}.
\newblock Technical Report Aeronautical Research Council Reports and Memoranda
  3438, Ministry of Aviation, London, 1964.
\newblock URL
  \url{http://repository.tudelft.nl/view/aereports/uuid:a38f3704-18d9-4ac8-a204-14ae03d84d8c/}.

\bibitem[Rowley and Williams(2006)]{Rowley2006}
Clarence~W. Rowley and David~R. Williams.
\newblock {Dynamics and Control of High-Reynolds-Number Flow Over Open
  Cavities}.
\newblock \emph{Annual Review of Fluid Mechanics}, 38\penalty0 (1):\penalty0
  251--276, 2006.
\newblock ISSN 0066-4189.
\newblock \doi{10.1146/annurev.fluid.38.050304.092057}.

\bibitem[Rowley et~al.(2002)Rowley, Colonius, and Basu]{Rowley2002}
Clarence~W. Rowley, Tim Colonius, and Amit~J. Basu.
\newblock {On self-sustained oscillations in two-dimensional compressible flow
  over rectangular cavities}.
\newblock \emph{Journal of Fluid Mechanics}, 455:\penalty0 315--346, mar 2002.
\newblock ISSN 0022-1120.
\newblock \doi{10.1017/S0022112001007534}.
\newblock URL
  \url{http://www.journals.cambridge.org/abstract{\_}S0022112001007534}.

\bibitem[Silva et~al.(2010)Silva, Souza, and Medeiros]{Silva2010}
H.~G. Silva, L.~F. Souza, and Marcello A.~F. Medeiros.
\newblock {Verification of a mixed high-order accurate DNS code for laminar
  turbulent transition by the method of manufactured solutions}.
\newblock \emph{International Journal for Numerical Methods in Fluids},
  64\penalty0 (3):\penalty0 336--354, sep 2010.
\newblock ISSN 02712091.
\newblock \doi{10.1002/fld.2156}.
\newblock URL \url{http://doi.wiley.com/10.1002/fld.2156}.

\bibitem[Souza et~al.(2005)Souza, Mendon{\c{c}}a, and Medeiros]{Souza2005}
L.~F. Souza, M.~T. Mendon{\c{c}}a, and M.~A.~F. Medeiros.
\newblock {The advantages of using high-order finite differences schemes in
  laminar-turbulent transition studies}.
\newblock \emph{International Journal for Numerical Methods in Fluids},
  48\penalty0 (5):\penalty0 565--582, 2005.
\newblock ISSN 02712091.
\newblock \doi{10.1002/fld.955}.

\bibitem[Sun et~al.(2016)Sun, Taira, Cattafesta, and Ukeiley]{Sun2016}
Y.~Sun, K.~Taira, L.~N. Cattafesta, and L.~S. Ukeiley.
\newblock {Spanwise effects on instabilities of compressible flow over a long
  rectangular cavity}.
\newblock \emph{Theoretical and Computational Fluid Dynamics}, pages 1--11, nov
  2016.
\newblock ISSN 0935-4964.
\newblock \doi{10.1007/s00162-016-0412-y}.
\newblock URL \url{http://link.springer.com/10.1007/s00162-016-0412-y}.

\bibitem[Sun et~al.(2017)Sun, Taira, Cattafesta, and Ukeiley]{Sun2017}
Yiyang Sun, Kunihiko Taira, Louis~N. Cattafesta, and Lawrence~S. Ukeiley.
\newblock {Biglobal instabilities of compressible open-cavity flows}.
\newblock \emph{Journal of Fluid Mechanics}, 826:\penalty0 270--301, sep 2017.
\newblock ISSN 0022-1120.
\newblock \doi{10.1017/jfm.2017.416}.
\newblock URL
  \url{https://www.cambridge.org/core/product/identifier/S0022112017004165/type/journal{\_}article}.

\bibitem[Tam(1976)]{Tam1976}
C.~K.~W. Tam.
\newblock {The acoustic modes of a two-dimensional rectangular cavity}.
\newblock \emph{Journal of Sound and Vibration}, 49\penalty0 (3):\penalty0
  353--364, 1976.
\newblock ISSN 10958568.
\newblock \doi{10.1016/0022-460X(76)90426-0}.

\bibitem[Tezuka and Suzuki(2006)]{Tezuka2006}
Asei Tezuka and Kojiro Suzuki.
\newblock {Three-dimensional global linear stability analysis of flow around a
  spheroid}.
\newblock \emph{AIAA journal}, 44\penalty0 (8):\penalty0 1697--1708, 2006.
\newblock ISSN 0001-1452.
\newblock \doi{10.2514/1.16632}.
\newblock URL \url{http://arc.aiaa.org/doi/pdf/10.2514/1.16632}.

\bibitem[Theofilis(2011)]{Theofilis2011}
Vassilios Theofilis.
\newblock {Global Linear Instability}.
\newblock \emph{Annual Review of Fluid Mechanics}, 43\penalty0 (1):\penalty0
  319--352, 2011.
\newblock ISSN 0066-4189.
\newblock \doi{10.1146/annurev-fluid-122109-160705}.
\newblock URL
  \url{http://www.annualreviews.org/doi/suppl/10.1146/annurev-fluid-122109-160705}.

\bibitem[Yamouni et~al.(2013)Yamouni, Sipp, and Jacquin]{Yamouni2013}
Sami Yamouni, Denis Sipp, and Laurent Jacquin.
\newblock {Interaction between feedback aeroacoustic and acoustic resonance
  mechanisms in a cavity flow: a global stability analysis}.
\newblock \emph{Journal of Fluid Mechanics}, 717:\penalty0 134--165, feb 2013.
\newblock ISSN 0022-1120.
\newblock \doi{10.1017/jfm.2012.563}.
\newblock URL
  \url{http://www.journals.cambridge.org/abstract{\_}S0022112012005630}.

\end{thebibliography}

\appendix

\section{Code validation and grid independence tests}

\label{sec:validation}

\subsection{Flow solver}

\label{sec:validationSolver}

The test case used as a reference for the DNS validation is described by \citet{Colonius1999}. Results of the validation are presented here, but further details can be found in \citet{Mathias2018}. The cavity's aspect ratio is $L/D=4$ and $L/\theta=102$, Reynolds and Mach numbers are, respectively, $Re_\theta=60$ ($Re_D=1530$) and $Ma=0.6$.

\begin{table}[h!]
	\begin{center}
		\caption{Meshes for the DNS grid independence analysis.}
		\begin{tabular}{c|c|c}
			& Mesh 1 & Mesh 2 \\ 
			\hline 
			Nodes in $x$ & 300 & 400 \\ 
			Nodes in $y$ & 150 & 200 \\ 
			Nodes in the cavity& 111$\times$71& 147$\times$94
		\end{tabular} 
		\label{tab:meshesDNSval}
	\end{center}
\end{table}

Two meshes were used to verify that the results were grid independent (table~\ref{tab:meshesDNSval}). Both meshes covered the same domain, from $x_i=-2$ to $x_f=15$ and from $y_i=-1$ to $y_f=4$. The cavity ends are at $x_1=5.34$ and $x_2=9.34$. Both meshes are stretched so that the most refined region is in the mixing layer at the cavity opening. The validation grids also shared the same buffer zone parameters, which added 20 nodes at each open domain boundary.

\begin{figure}[h!]
	\begin{center}
		\includegraphics[width=0.8\textwidth] {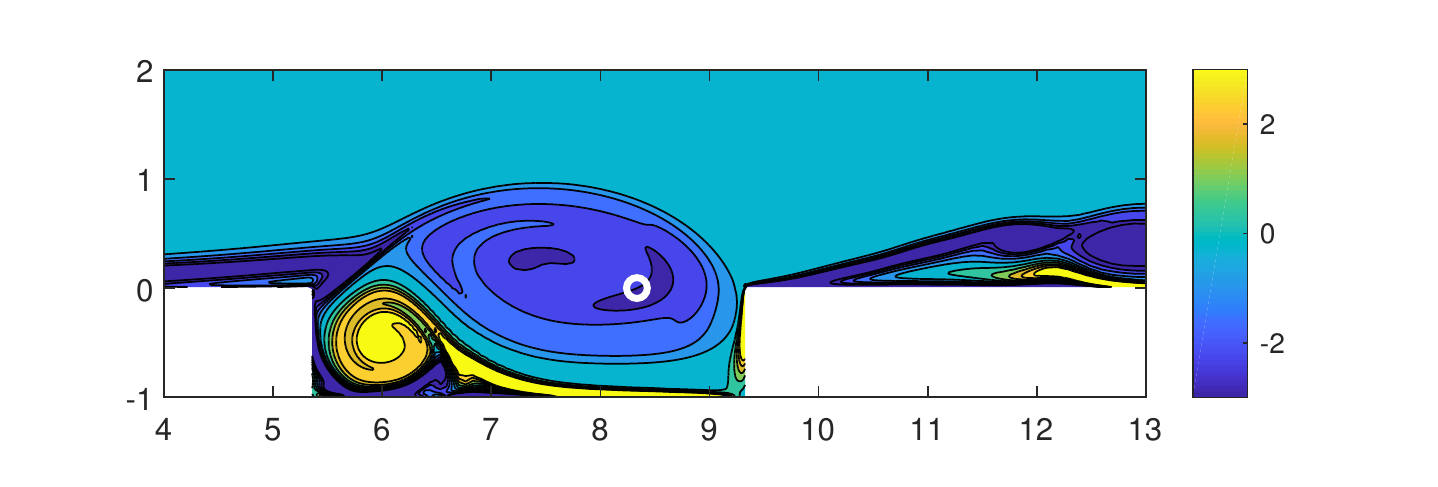}
	\end{center}
	\caption{Vorticity contours of the unsteady flow at an arbitrary time. The white circle marks a point where data is collected for figure~\ref{fig:2dUnsteadyVelocity}}
	\label{fig:unsteadyFlow}
\end{figure}

\begin{figure}[h!]
	\begin{center}
		\includegraphics[width=0.9\textwidth] {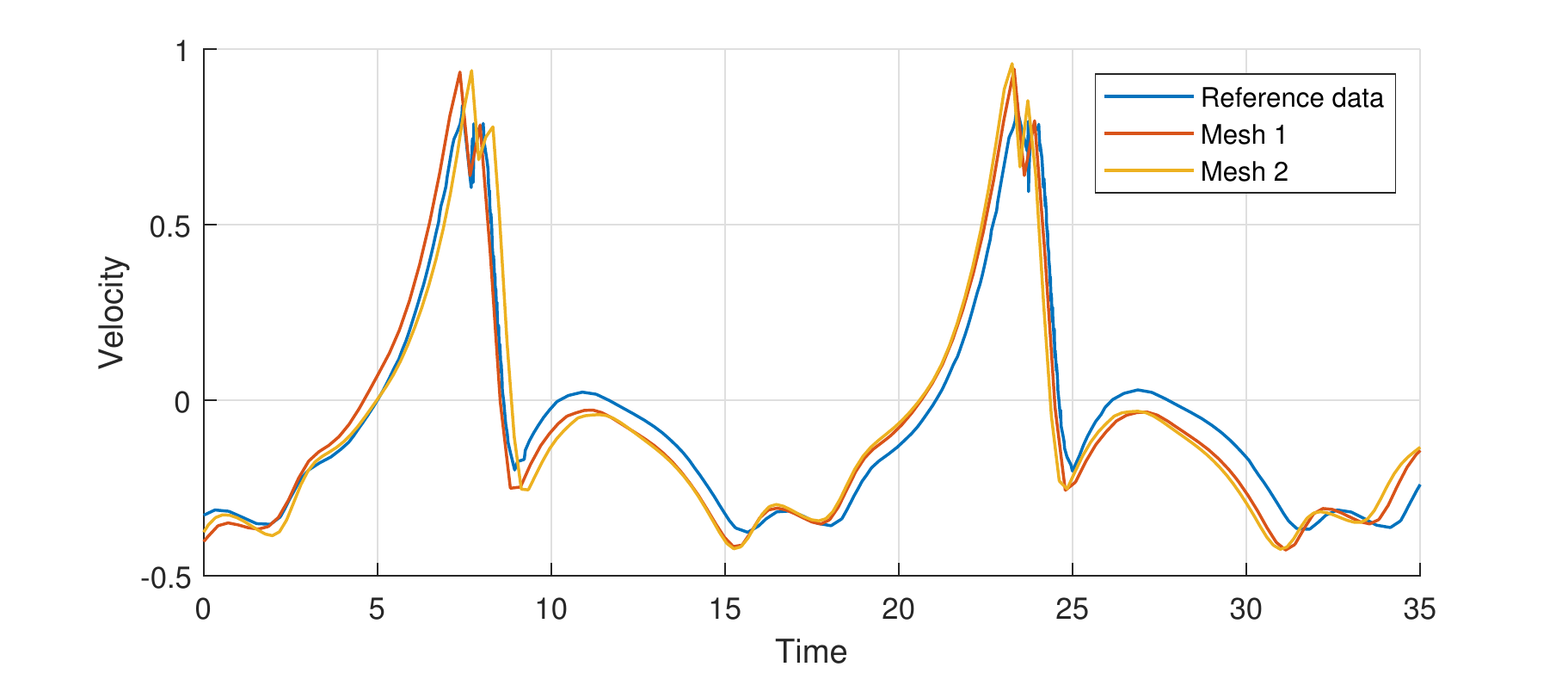}
	\end{center}
	\caption{Velocity as a function of time at a fixed point for an unstable case in periodic state.}
	\label{fig:2dUnsteadyVelocity}
\end{figure}

Figure~\ref{fig:unsteadyFlow} illustrates the flow at an arbitrary time after a periodic state is established. It shows a single vortex inside the cavity and vortices being shed from the cavity. Figure~\ref{fig:2dUnsteadyVelocity} shows the wall-normal velocity as a function of time at the point shown in figure~\ref{fig:unsteadyFlow}, three quarters across the cavity opening. Data extracted from the reference paper is also plotted. The phases were manually adjusted for better comparison. The results from our computations were grid independent and agreed with the reference results.

\subsection{Global stability analysis}

\label{sec:validationGlobal}

The instability results have to be independent of the computational grid, domain and other computational parameters. The analysis is presented for the thick boundary layer case at $Ma=0.1$, as an example, but similar conclusions were obtained for other cases. Corresponding tests were carried out for the base flow, but this is much less demanding and not presented.

\begin{table}[h!]
	\begin{center}
		\caption{Meshes for global instability grid independence analysis.}
		\begin{tabular}{c|c|c|c|c}
			& Mesh 1 & Mesh 2 & Mesh 3 & Mesh 4 \\ 
			\hline 
			Nodes in $x$ & 200 & 300 & 300 & 400 \\ 
			Nodes in $y$ & 150 & 150 & 225 & 300 \\ 
			Nodes in the cavity& 76$\times$52& 114$\times$52& 114$\times$79& 152$\times$105
		\end{tabular} 
		\label{tab:meshesConv2d}
	\end{center}
\end{table}

For the mesh independence analysis, four meshes were used, as shown in table~\ref{tab:meshesConv2d}. All these meshes are for the same domain, which spans from $x_i=-2$ to $x_f=10$ and from $y_i=-1$ to $y_f=4$. The cavity is placed from $x_1=2.96$ to $x_2=4.96$. The buffer zone used employed the same parameters of the DNS validation tests.

For meshes 1 to 4, the time steps used were $8 \times 10^{-4}$, $6 \times 10^{-4}$, $5 \times 10^{-4}$ and $4 \times 10^{-4}$, respectively and the number of steps, 500, 667, 800 and 1000, so that for all cases the total physical time simulated was identical. The stretching parameters were kept constant, therefore Mesh 4 is twice as refined as Mesh 1 in both directions.

\begin{figure}[h!]
	\begin{center}
		\includegraphics[width=0.9\textwidth] {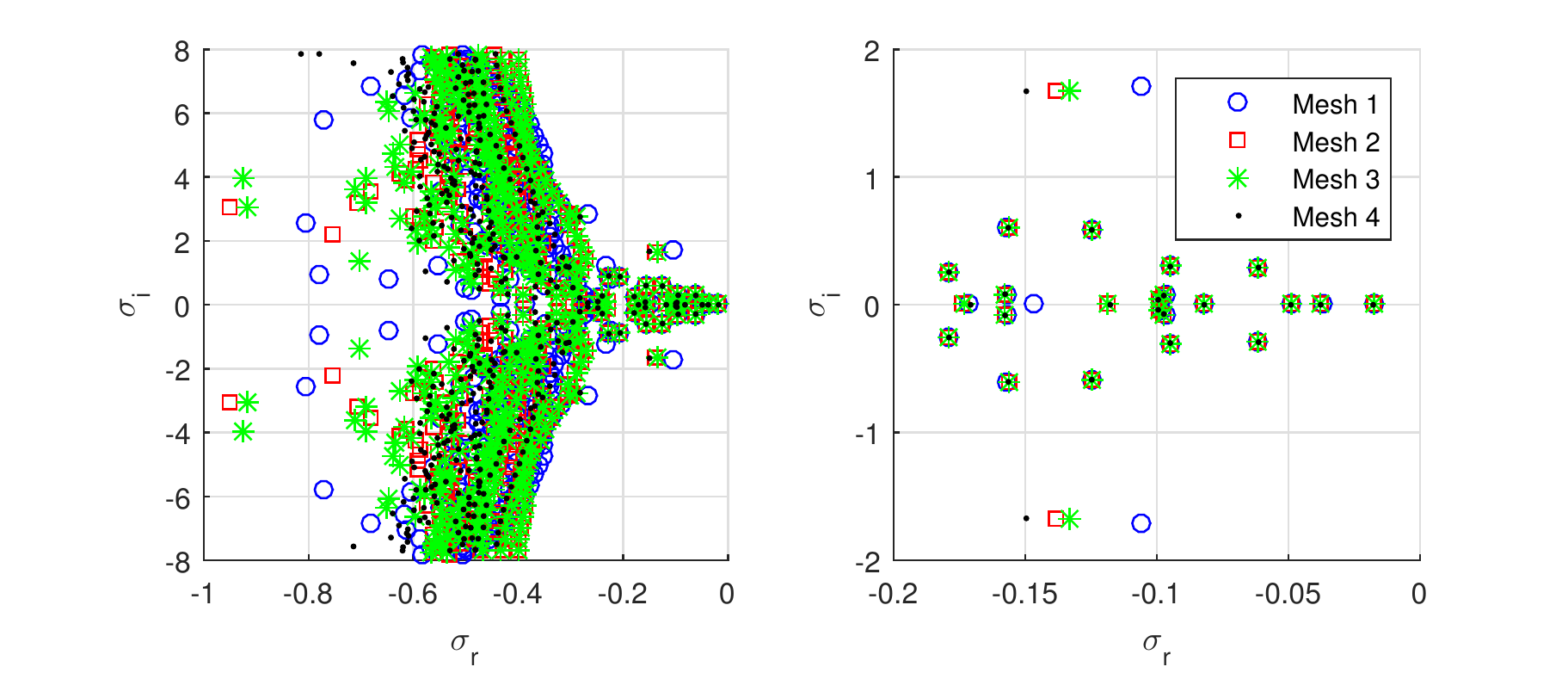}
	\end{center}
	\caption{Eigenvalues computed for the mesh independence analysis.}
	\label{fig:meshConvergenceComplexPlane}
\end{figure}

Figure~\ref{fig:meshConvergenceComplexPlane} shows that the 12 leading eigenvalues obtained for all meshes in the complex plane are very close. The values of the 15 leading eigenvalues are shown in table~\ref{tab:meshesConv2dEigenvalues} and, for the most refined meshes, agree within three decimal places. There is also no visible difference between the eigenmodes obtained for each mesh as well.

\begin{table}[h!]
	\footnotesize
	\begin{center}
		\caption{Eigenvalues from mesh independence analysis.}
		\setlength{\tabcolsep}{3pt}
		\begin{tabular}{c|cccc|cccc}
			&\multicolumn{4}{c|}{Real part} & \multicolumn{4}{c}{Imaginary part} \\
			Mode & Mesh 1 & Mesh 2 & Mesh 3 & Mesh 4 & Mesh 1 & Mesh 2 & Mesh 3 & Mesh 4 \\ 
			\hline 
			1& -0.017783& -0.017811& -0.017819& -0.017826& -& -& -& -\\
			2& -0.037405& -0.037671& -0.037696& -0.037753& -& -& -& -\\
			3& -0.049209& -0.049089& -0.049176& -0.049114& -& -& -& -\\
			4& -0.061377& -0.061300& -0.061370& -0.061356& 0.293135& 0.292746& 0.291815& 0.292134\\
			5& -0.061377& -0.061300& -0.061370& -0.061356& -0.293135& -0.292746& -0.291815& -0.292134\\
			6& -0.082059& -0.082041& -0.081997& -0.082013& -& -& -& -\\
			7& -0.095062& -0.094852& -0.095024& -0.094946& 0.300320& 0.300000& 0.299059& 0.299369\\
			8& -0.095062& -0.094852& -0.095024& -0.094946& -0.300320& -0.300000& -0.299059& -0.299369\\
			9& -0.096846& -0.097439& -0.097109& -0.097378& 0.071643& 0.072253& 0.071825& 0.072137\\
			10& -0.096846& -0.097439& -0.097109& -0.097378& -0.071643& -0.072253& -0.071825& -0.072137\\
			11& -0.098625& -0.099305& -0.099338& -0.099503& 0.039301& 0.039409& 0.039239& 0.039309\\
			12& -0.098625& -0.099305& -0.099338& -0.099503& -0.039301& -0.039409& -0.039239& -0.039309\\
			13& -0.106347& -0.118776& -0.119008& -0.117730& 1.704951& -& -& -\\
			14& -0.106347& -0.124620& -0.124782& -0.124740& -1.704951& 0.587605& 0.585761& 0.586391\\
			15& -0.124803& -0.124620& -0.124782& -0.124740& 0.588443& -0.587605& -0.585761& -0.586391\\
		\end{tabular} 
		\label{tab:meshesConv2dEigenvalues}
	\end{center}
\end{table}

\begin{figure}[h!]
	\begin{center}
		\includegraphics[width=0.8\textwidth] {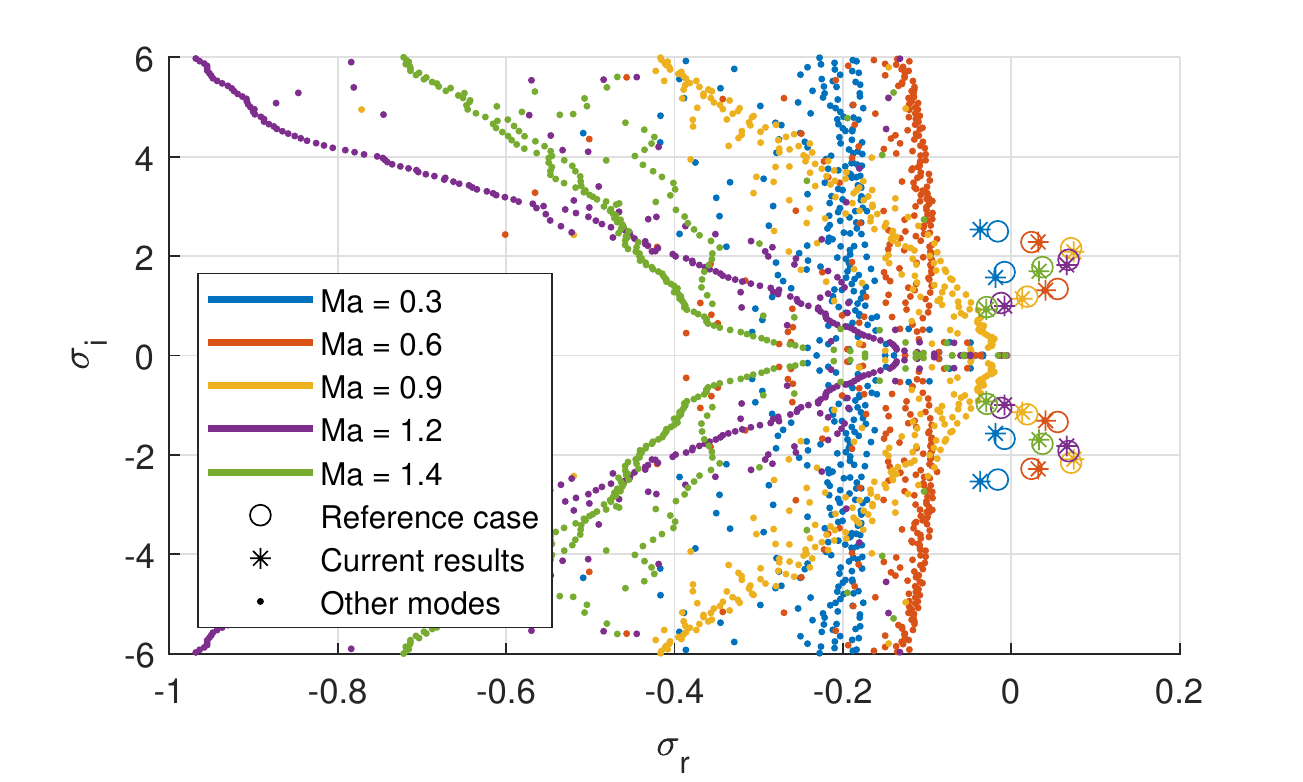}
	\end{center}
	\caption{Eigenspectra computed for $Ma=0.3$, $0.6$, $0.9$, $1.2$ and $1.4$, and comparison with reference values by \citet{Sun2017}.}
	\label{fig:sun2017}
\end{figure}

A set of results for Rossiter mode stability was provided by \citet{Sun2017}, where eigenvalues for both modes 1 and 2 were given for Mach numbers between 0.3 and 1.4, $Re_D=1500$, $D/\theta=26.4$, and $L/D=2$. Figure~\ref{fig:sun2017} compares those results to ours at the same parameters. The agreement is good, particularly the trends of all unstable modes with $Ma$. It can also be added that perfect agreement can be difficult in global instability analysis of open flows because often not enough information is given of the infinity domain boundary conditions used in every study and different conditions at these boundaries can significantly affect the instability results \cite{Ohmichi2016}. In our case, the inlet is upstream of the leading edge, and the cavity position is chosen so that the $D/\theta$ equals the selected value for a Blasius boundary layer. In \citet{Sun2017}, the domain inlet is downstream of the flat plate leading edge and is given by a Blasius boundary layer profile. This difference in set-up, for example, may lead to significant difference in the boundary layer thicknesses over the cavity.

Further validation results for both the flow solver and the global stability routine can be found in \citet{Mathias2018}.

\section{Verification of Absolute stability}

\label{sec:absoluteInstability}

\begin{figure}[h!]
	\begin{center}
		\includegraphics[width=0.25\textwidth]{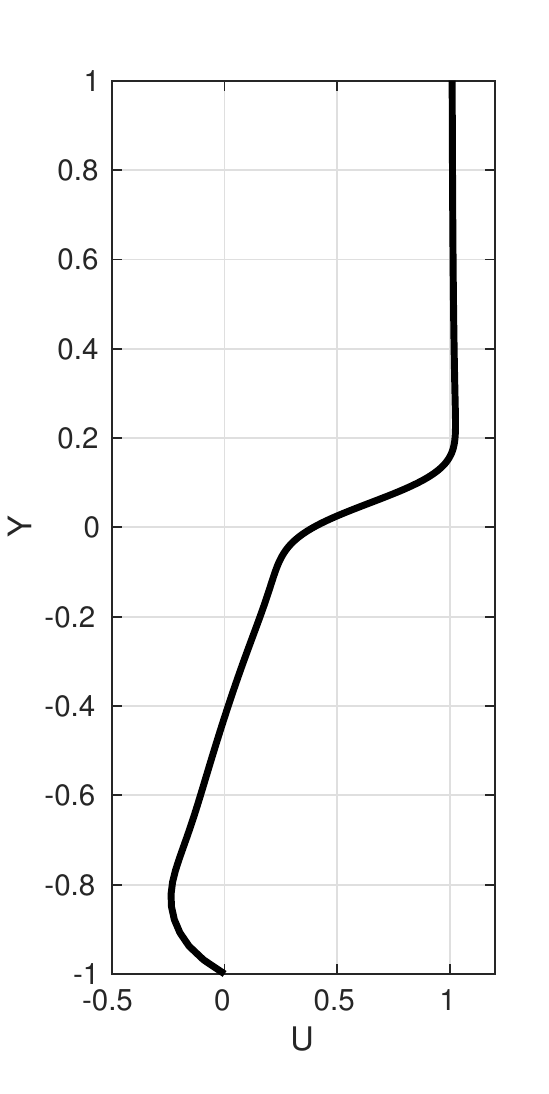}
		\includegraphics[width=0.66\textwidth]{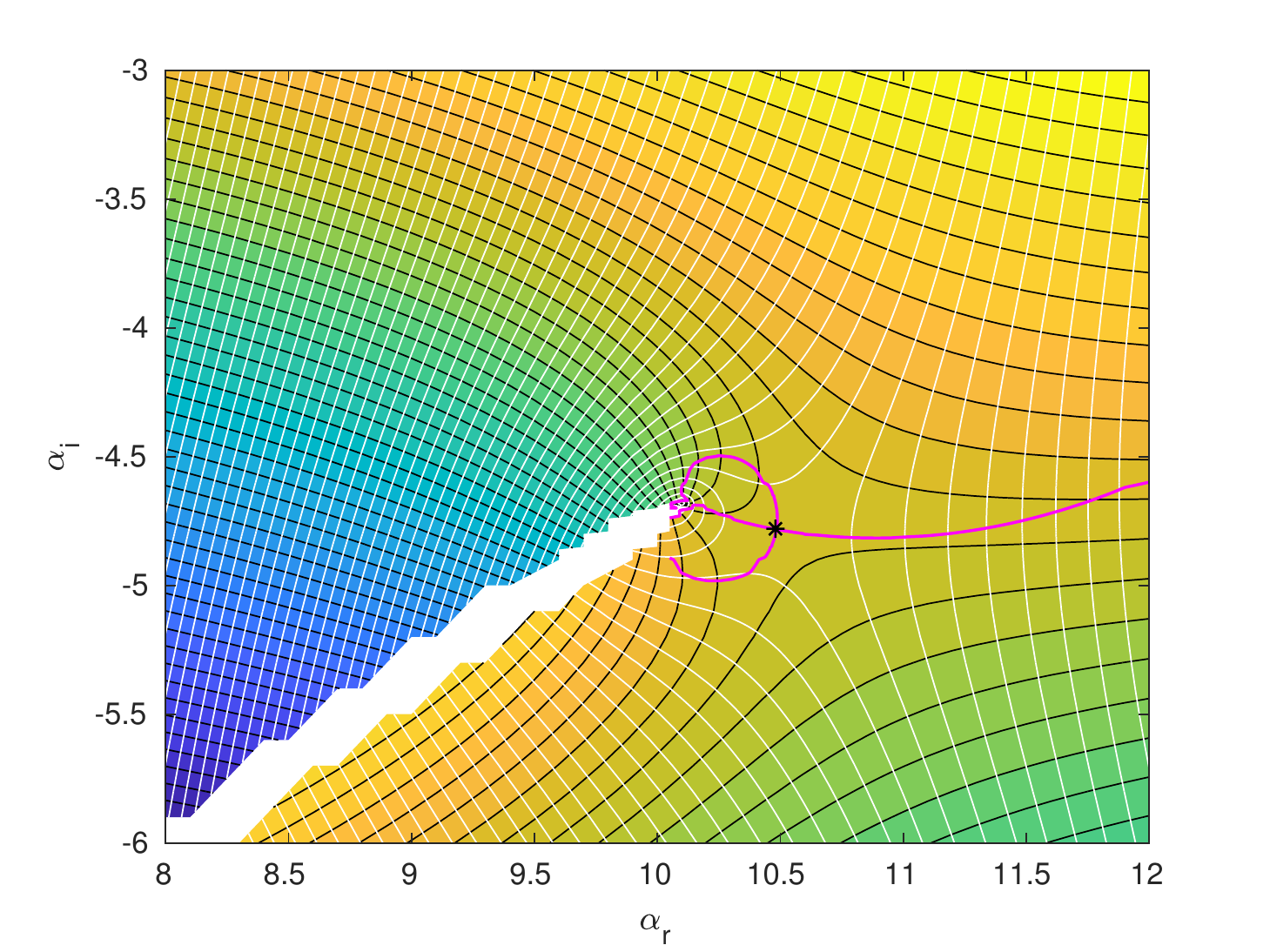}
	\end{center}
	\caption{(a) Velocity profile considered for the absolute instability analysis. (b) Complex $\omega$ mapped as a function of complex $\alpha$. The colors and black line contours indicate the imaginary part and the white lines, the real part. The magenta lines cross at the saddle point.}
	\label{fig:abs_inst}
\end{figure}

The existence of relatively large reversed flow in the cavity raises the possibility of a local absolute instability. If a localized spot of absolute instability exists, modes other than the Rossiter ones could become globally unstable. We investigated this possibility by performing an absolute local instability analysis. The analysis did not include compressibility because this has a small and stabilizing effect at subsonic conditions. The profile with highest reversed flow (23.8\%) was used in the analysis and corresponded to the $D/\theta=100$ case. Changing the Mach number caused only a negligible change to the profile, hence, $Ma=0.5$ was chosen. Figure~\ref{fig:abs_inst}(a) depicts this velocity profile.

For the local instability analysis we solved the Orr-Sommerfeld equation and followed \citet{Juniper2014}. Figure~\ref{fig:abs_inst}(b) shows $\omega$ (the complex frequency) as a function of $\alpha$ (the complex wavenumber). The black lines (and the colors) are isocontours of the imaginary part and the white lines, of the real part. The absolute instability is determined by the condition at the saddle point. The magenta lines mark the location where either $\partial \omega_i / \partial \alpha_r$ or $\partial \omega_i / \partial \alpha_i$ vanish. They cross each other at the saddle point, where $\partial \omega/ \partial \alpha = 0$, i.e., the group velocity is null. In this case, it happens at $\alpha = 10.5 - 4.8i$, where $\omega = 4.9 - 2.8i$. The negative imaginary part of $\omega$ indicates that this flow is locally absolutely stable.

The fact that this flow is not locally absolutely unstable was already expected because several studies on this parametric region \cite{Rowley2002, Yamouni2013, Meseguer2014}, despite having not carried out this analysis, have also not reported modes other than the Rossiter one for $L/D=2$.

\end{document}